\documentclass[jrnl_compsoc]{IEEEtran}
\ifCLASSINFOpdf
   \usepackage[pdftex]{graphicx}
\else
\fi

\usepackage{algorithm}
\usepackage{algpseudocode}

\usepackage{times}
\usepackage[us,12hr]{datetime}
\usepackage{enumitem}
\usepackage{subcaption}
\usepackage{comment}
\usepackage{multirow}
\usepackage{amsmath}
\usepackage{amssymb}
\usepackage{wrapfig}

\RequirePackage[amsmath,hyperref,thmmarks]{ntheorem}
\newtheorem{theorem}{Theorem}

\newtheorem{definition}{Definition}

\theoremstyle{nonumberplain}
\theoremheaderfont{\normalfont\itshape}
\theorembodyfont{\normalfont}
\theoremsymbol{\ensuremath{\square}}

\usepackage{tikzpagenodes}

\newcommand{\itembase}[1]{\setlength{\itemsep}{#1}}

\begin{document}
%
\title{Distillation as a Defense to Adversarial \\ Perturbations against Deep Neural Networks}

\author{\IEEEauthorblockN{Nicolas Papernot\IEEEauthorrefmark{1}, Patrick McDaniel\IEEEauthorrefmark{1}, Xi Wu\IEEEauthorrefmark{4}, Somesh Jha\IEEEauthorrefmark{4}, and Ananthram Swami\IEEEauthorrefmark{3}\\}
\IEEEauthorblockA{\IEEEauthorrefmark{1}Department of Computer Science and Engineering, Penn State University\\}
\IEEEauthorblockA{\IEEEauthorrefmark{4}Computer Sciences Department, University of Wisconsin-Madison\\}
\IEEEauthorblockA{\IEEEauthorrefmark{3}United States Army Research Laboratory, Adelphi, Maryland}\\
\IEEEauthorblockA{\{ngp5056,mcdaniel\}@cse.psu.edu, \{xiwu,jha\}@cs.wisc.edu, ananthram.swami.civ@mail.mil}\vspace{-0.4in}}

\maketitle

  \begin{tikzpicture}[remember picture,overlay]
    \node[align=center] at ([yshift=1em]current page text area.north) {Accepted to the 37th IEEE Symposium on Security \& Privacy, IEEE 2016. San Jose, CA. };
  \end{tikzpicture}%

  \vspace{-0.2in}




\begin{abstract} 
Deep learning algorithms have been shown to perform extremely well on many
classical machine learning problems.   However, recent studies have shown that
deep learning, like other machine learning techniques, is vulnerable to adversarial samples: inputs crafted to force a
deep neural network (DNN) to provide adversary-selected outputs. Such attacks
can seriously undermine the security of the system supported by the DNN,
sometimes with devastating consequences.  For example, autonomous vehicles can
be crashed, illicit or illegal content can bypass content filters, or biometric
authentication systems can be manipulated to allow improper access.  In this
work, we introduce a defensive mechanism called \emph{defensive distillation}
to reduce the effectiveness of adversarial samples on DNNs. We analytically
investigate the generalizability and robustness properties granted by the use
of defensive distillation when training DNNs. We also empirically study the
effectiveness of our defense mechanisms on two DNNs placed in adversarial
settings.  The study shows that defensive distillation can reduce effectiveness
of sample creation from 95\% to less than 0.5\% on a studied DNN. Such dramatic
gains can be explained by the fact that distillation leads gradients used in
adversarial sample creation to be reduced by a factor of $10^{30}$. We also
find that distillation increases the average minimum number of features that
need to be modified to create adversarial samples by about 800\% on one of the
DNNs we tested. 
\end{abstract}


%
\IEEEpeerreviewmaketitle

\section{Introduction}

\emph{Deep Learning} (DL) has been demonstrated to perform exceptionally well
on several categories of machine learning problems, notably input
classification.  These \emph{Deep Neural Networks} (DNNs) efficiently learn
highly accurate models from a large corpus of training samples, and thereafter
classify unseen samples with great accuracy. As a result, DNNs are used in many
settings~\cite{krizhevsky2012imagenet, sainath2013deep, sermanet2014overfeat},
some of which are increasingly security-sensitive~\cite{dahl2013large,
yuan2014droid, PayPal}. By using deep learning algorithms, designers of these
systems make implicit security assumptions about deep neural networks. However,
recent work in the machine learning and security communities have shown that
adversaries can force many machine learning models, including DNNs, to produce adversary-selected outputs using
carefully crafted inputs~\cite{NAS-186, szegedy2013intriguing,
goodfellow2014explaining}. 

Specifically, adversaries can craft particular inputs, named
\emph{adversarial samples},  leading models to produce an output behavior of their
choice, such as misclassification. Inputs are crafted by adding a carefully
chosen adversarial perturbation to a legitimate sample. The resulting sample is
not necessarily unnatural, i.e. outside of the training data manifold.
Algorithms crafting adversarial samples are designed to minimize the
perturbation, thus making adversarial samples hard to distinguish from
legitimate samples. Attacks based on adversarial samples occur after training
is complete and therefore do not require any tampering with the training
procedure. 

To illustrate how adversarial samples make a system based on DNNs vulnerable,
consider the following input samples:

\vspace{3pt}

\centerline{\includegraphics[width=0.2\textwidth]{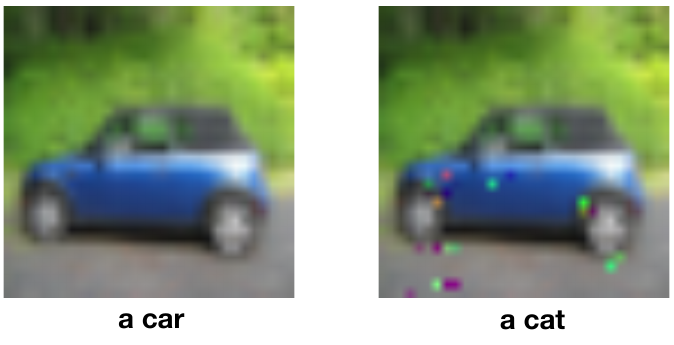}}

\noindent
 The left image is correctly classified by a trained DNN as a car.  The right
image was crafted by an adversarial sample algorithm (in~\cite{NAS-186}) from
the correct left image.  The altered image is incorrectly classified as a cat
by the DNN.  To see why such misclassification is dangerous, consider deep
learning as it is commonly used in autonomous (driverless)
cars~\cite{NVIDIATegra}. Systems based on DNNs are used to recognize signs or
other vehicles on the road~\cite{cirecsan2012multi}. If perturbing the input of
such systems, by slightly altering the car's body for instance, prevents DNNs
from classifying it as a moving vehicule correctly, the car might not stop and
eventually be involved in an accident, with potentially disastrous
consequences. The threat is real where an adversary can profit from evading
detection or having their input misclassified.  Such attacks commonly occur
today in non-DL classification systems~\cite{huang2011adversarial,
biggio2014pattern, biggio2013evasion, anjos2011counter, fogla2006evading}.

Thus, adversarial samples must be taken into account when designing security
sensitive systems incorporating DNNs. Unfortunately, there are very few
effective countermeasures available today.  Previous work considered the
problem of constructing such defenses but solutions proposed are deficient in
that they require making modifications to the DNN architecture or only
partially prevent adversarial samples from being
effective~\cite{goodfellow2014explaining, gu2014towards} (see
Section~\ref{sec:related-work}). 

\emph{Distillation} is a training procedure initially designed to train a DNN
using knowledge transferred from a different DNN. The intuition was suggested
in~\cite{ba2014deep} while distillation itself was formally introduced
in~\cite{hinton2015distilling}. The motivation behind the knowledge transfer
operated by distillation is to reduce the computational complexity of DNN
architectures by transferring knowledge from larger architectures to smaller
ones. This facilitates the deployment of deep learning in resource constrained
devices (e.g. smartphones) which cannot rely on powerful GPUs to perform
computations. \emph{We formulate a new variant of distillation to provide for
defense training:} instead of transferring knowledge between  different
architectures, we propose to use the knowledge extracted from a DNN to improve
its own resilience to adversarial samples. 

In this paper, we explore analytically and empirically the use of distillation
as a defensive mechanism against adversarial samples.  We use the knowledge
extracted during distillation to reduce the amplitude of network gradients
exploited by adversaries to craft adversarial samples.  If adversarial
gradients are high, crafting adversarial samples becomes easier because small
perturbations will induce high DNN output variations. To defend against such
perturbations, one must therefore reduce variations around the input, and
consequently the amplitude of adversarial gradients. In other words, we use
defensive distillation to smooth the model learned by a DNN architecture during
training by helping the model generalize better to samples outside of its
training dataset. 

 \emph{At test time, models trained with defensive distillation are less
sensitive to adversarial samples}, and are therefore more suitable for
deployment in security sensitive settings. We make the following contributions
in this paper:
 
\begin{itemize}
\itembase{3pt}

\item We articulate the requirements for the design of adversarial sample DNN
defenses. These guidelines highlight the inherent tension between defensive
robustness, output accuracy, and performance of DNNs.

\item We introduce \emph{defensive distillation}, a procedure to train
DNN-based classifier models that are more robust to perturbations. Distillation
extracts additional knowledge about training points as class probability
vectors produced by a DNN, which is fed back into the training regimen.  This
departs substantially from the past uses of distillation which aimed to reduce
the DNN architectures to improve computational performance, but rather feeds
the gained knowledge back into the original models.

\item We analytically investigate defensive distillation as a security
countermeasure. We show that distillation  generates smoother classifier models
by reducing their sensitivity to input perturbations. These smoother DNN
classifiers are found to be more resilient to adversarial samples and have
improved class generalizability properties. 

\item We show empirically that defensive distillation reduces the success rate
of adversarial sample crafting from $95.89\%$ to $0.45\%$ against a first DNN
trained on the MNIST dataset~\cite{lecun1998mnist}, and from $87.89\%$ to
$5.11\%$ against a second DNN trained on the
CIFAR10~\cite{krizhevsky2009learning} dataset.

\item A further empirical exploration of the distillation parameter space shows
that a correct parameterization can reduce the sensitivity of a DNN to input
perturbations by a factor of $10^{30}$. Successively, this increases the
average minimum number of input features to be perturbed to achieve adversarial
targets by $790\%$ for a first DNN, and by $556\%$ for a second DNN.

\end{itemize}


\section{Adversarial Deep Learning}
\label{sec:adversarial-deep-learning}

Deep learning is an established technique in machine learning. In this section,
we provide some rudiments of deep neural networks (DNNs) necessary to
understand the subtleties of their use in adversarial settings. We then
formally describe two attack methods in the context of a framework that we
construct to (i) develop an understanding of DNN vulnerabilities exploited by
these attacks and (ii) compare the strengths and weaknesses of both attacks in
various adversarial settings. Finally, we provide an overview of a DNN training
procedure, which our defense mechanism builds on, named distillation.

\subsection{Deep Neural Networks in Adversarial Settings}

\begin{figure}
\centering
\includegraphics[width=\columnwidth]{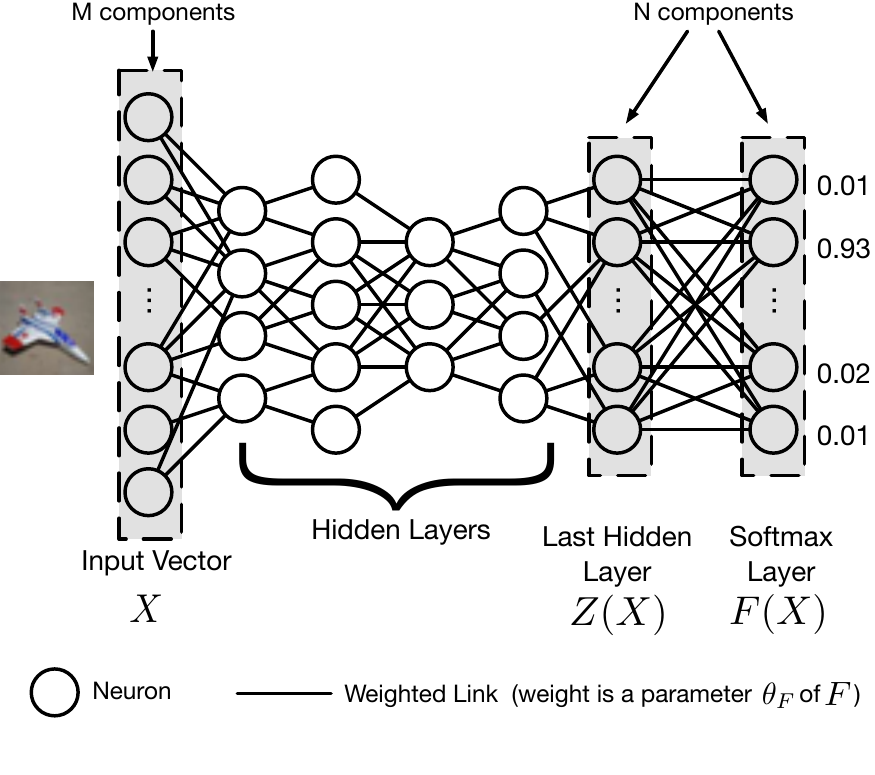}
\caption{\textbf{Overview of a DNN architecture:} This architecture, suitable for classification tasks thanks to its softmax output layer, is used throughout the paper along with its notations.}
\label{fig:dnn-architecture}
\end{figure}

\textbf{Training and deploying DNN architectures -} Deep neural networks
compose many parametric functions to build increasingly complex representations
of a high dimensional input expressed in terms of previous simpler
representations~\cite{Bengio-et-al-2015-Book}. Practically speaking, a DNN is
made of several successive layers of neurons building up to an output layer.
These layers can be seen as successive representations of the input
data~\cite{hinton2007learning}, a multidimensional vector $X$, each of them
corresponding to one of the parametric functions mentioned above. Neurons
constituting layers are modeled as elementary computing units applying an
activation function to their input. Layers are connected using links weighted
by a set of vectors, also referred to as network parameters $\theta_F$.
Figure~\ref{fig:dnn-architecture} illustrates such an architecture along with
notations used in this paper.

The numerical values of weight vectors in $\theta_F$ are evaluated during the
network's \emph{training phase}. During that phase, the DNN architecture is
given a large set of known input-output pairs $(X,Y)\in (\cal{X},\cal{Y})$. It
uses a series of successive forward and backward passes through the DNN layers
to compute prediction errors made by the output layer of the DNN, and
corresponding gradients with respect to weight
parameters~\cite{rumelhart1988learning}. The weights are then updated, using
the previously described gradients, in order to improve the prediction and
eventually the overall accuracy of the network. This training process is
referred to as \emph{backpropagation} and is governed by
\emph{hyper-parameters} essential to the  convergence of model
weight~\cite{bergstra2012random}. The most important hyper-parameter is the
\emph{learning rate} that controls the speed at which weights are updated with
gradients. 

Once the network is trained, the architecture together with its parameter
values $\theta_F$ can be considered as a classification function $F$ and the
\emph{test phase} begins: the network is used on unseen inputs $X$ to predict
outputs $F(X)$. Weights learned during training hold knowledge that the DNN
applies to these new and unseen inputs. Depending on the type of output
expected from the network, we either refer to \emph{supervised learning} when
the network must learn some association between inputs and outputs (e.g.,
classification~\cite{krizhevsky2012imagenet, dahl2013large, cirecsan2012multi,
glorot2011domain}) or \emph{unsupervised learning} when the network is trained
with unlabeled inputs (e.g., dimensionality reduction, feature engineering, or
network pre-training~\cite{krizhevsky2009learning, masci2011stacked,
erhan2010does}). In this paper, we only consider supervised learning, and more
specifically the task of classification. The goal of the training phase is to
enable the neural network to extrapolate from the training data it
observed during training so as to correctly predict outputs on new and unseen
samples at test time. 

\begin{figure}
\centering
\includegraphics[width=\columnwidth]{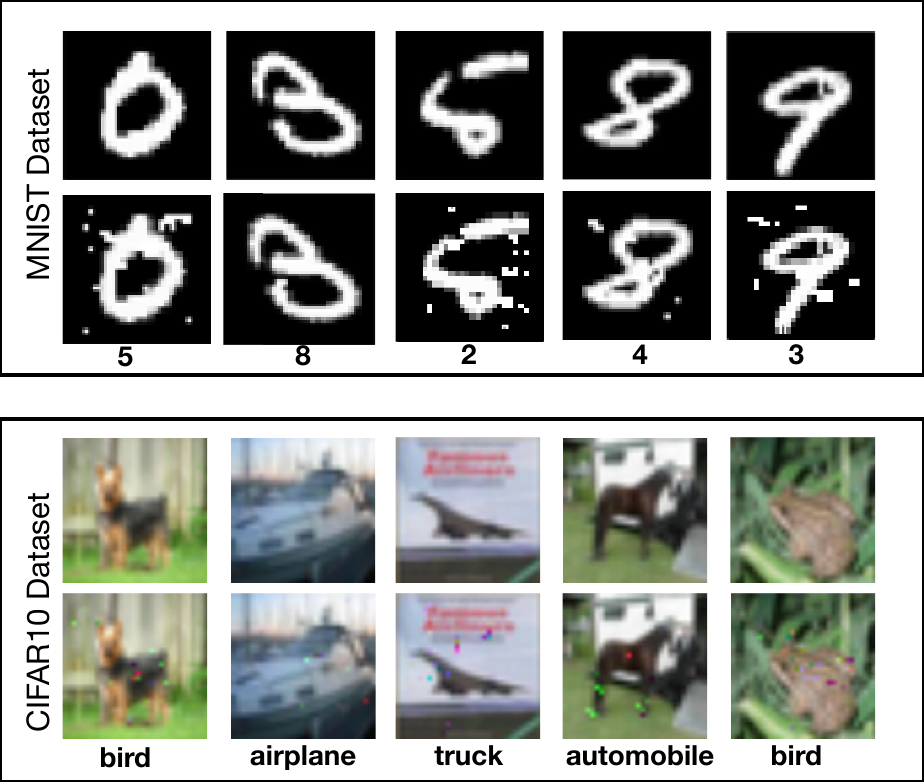}
\caption{\textbf{Set of legitimate and adversarial samples for two datasets:} For each dataset, a set of legitimate samples, which are correctly classified by DNNs, can be found on the top row while a corresponding set of adversarial samples (crafted using~\cite{NAS-186}), misclassifed by DNNs, are on the bottom row.}
\label{fig:adversarial-samples}
\end{figure}

\textbf{Adversarial Deep Learning -} It has been shown in previous work that
when DNNs are deployed in adversarial settings, one must take into account
certain vulnerabilities~\cite{NAS-186, szegedy2013intriguing,
goodfellow2014explaining}. Namely, \emph{adversarial samples} are artifacts of
a threat vector against DNNs that can be exploited by adversaries at test time,
after network training is completed. Crafted by adding carefully selected
perturbations $\delta X$ to legitimate inputs $X$, their key property is to
provoke a specific behavior from the DNN, as initially chosen by the adversary.
For instance, adversaries can alter samples to have them misclassified by a
DNN, as is the case of adversarial samples crafted in experiments presented in
section~\ref{sec:evaluation}, some of which are illustrated in
Figure~\ref{fig:adversarial-samples}. Note that the noise introduced by
perturbation $\delta X$ added to craft the adversarial sample must be small
enough to allow a human to still correctly process the sample.

Attacker's end goals can be quite diverse, as pointed out in previous work
formalizing the space of adversaries against deep learning~\cite{NAS-186}. For
classifiers, they range from simple confidence reduction (where the aim is to
reduce a DNN's confidence on a prediction, thus introducing class ambiguity),
to source-target misclassification (where the goal is to be able to take a
sample from any source class and alter it so as to have the DNN classify it in
any chosen target class distinct from the source class). This paper considers
the source-target misclassification, also known as the chosen target attack, in
the following sections. Potential examples of adversarial samples in realistic
contexts could include slightly altering malware executables in order to evade
detection systems built using DNNs, adding perturbations to handwritten digits
on a check resulting in a DNN wrongly recognizing the digits (for instance,
forcing the DNN to read a larger amount than written on the check), or altering
a pattern of illegal financial operations to prevent it from being picked up by
fraud detections systems using DNNs. Similar attacks occur today on non-DNN
classification systems~\cite{huang2011adversarial, biggio2014pattern,
biggio2013evasion, anjos2011counter} and are likely to be ported by adversaries
to DNN classifiers.

As explained later in the attack framework described in this section, methods
for crafting adversarial samples theoretically require a strong knowledge of
the DNN architecture. However in practice, even attackers with limited
capabilities can perform attacks by approximating their target DNN model $F$
and crafting adversarial samples on this approximated model. Indeed, previous
work reported that adversarial samples against DNNs are transferable from one
model to another~\cite{szegedy2013intriguing}. Skilled adversaries can thus
train their own DNNs to produce adversarial samples evading victim DNNs.
Therefore throughout this paper, we consider an attacker with the capability of
accessing a trained DNN used for classification, since the transferability of
adversarial samples makes this assumption acceptable. Such a capability can
indeed take various forms including for instance a direct access to the network
architecture implementation and parameters, or access to the network as an
oracle requiring the adversary to approximatively replicate the model. Note
that we do not consider attacks at training time in this paper and leave such
considerations to future work.

\subsection{Adversarial Sample Crafting}

\begin{figure*}
\centering
\includegraphics[width=\textwidth]{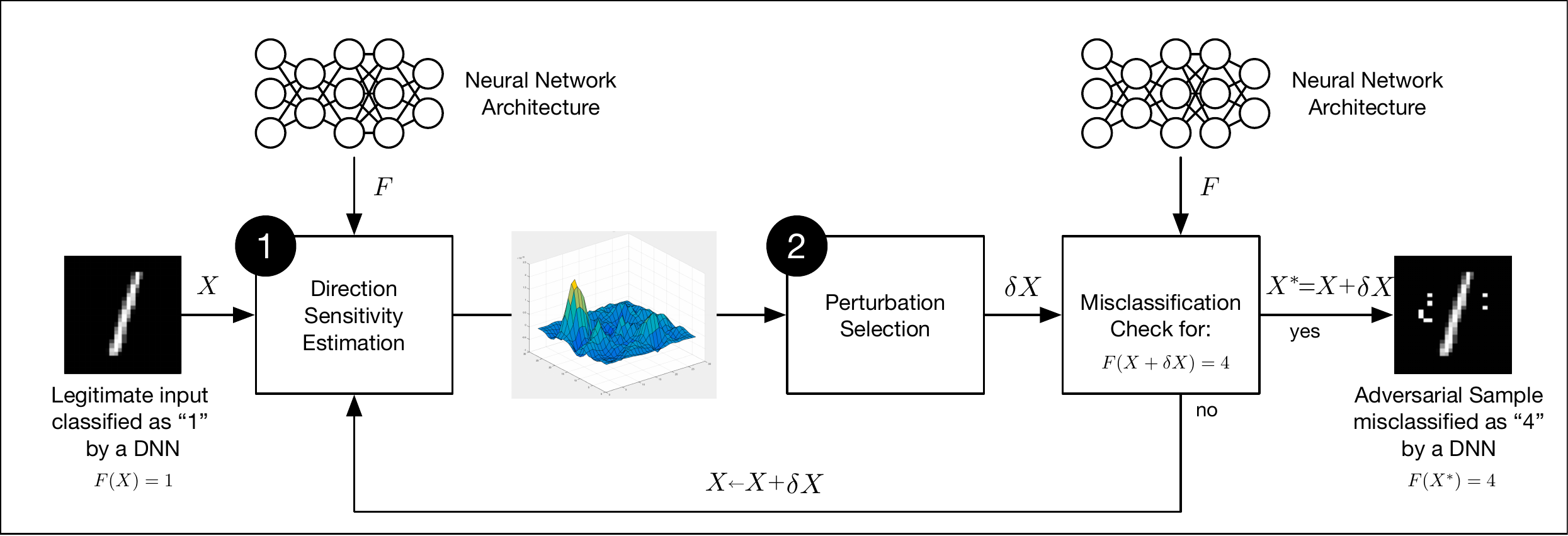}
\caption{\textbf{Adversarial crafting framework:} Existing algorithms for adversarial sample crafting~\cite{NAS-186, goodfellow2014explaining} are a succession of two steps: (1) \emph{direction sensitivity estimation} and (2) \emph{perturbation selection}. Step (1) evaluates the sensitivity of model $F$ at the input point corresponding to sample $X$. Step (2) uses this knowledge to select a perturbation affecting sample $X$'s classification. If the resulting sample $X+\delta X$ is misclassified by model $F$ in the adversarial target class (here 4) instead of the original class (here 1), an adversarial sample $X^*$ has been found. If not, the steps can be repeated on updated input $X\leftarrow X+\delta X$.}
\label{fig:adversarial-crafting-overview}
\end{figure*}

We now describe precisely how adversarial sample are crafted by adversaries.
The general framework we introduce builds on previous attack approaches and is
split in two folds: \emph{direction sensitivity estimation} and
\emph{perturbation selection}. Attacks holding in this framework correspond to
adversaries with diverse goals, including the goal of misclassifying samples
from a specific \emph{source class} into a distinct \emph{target class}. This
is one of the strongest adversarial goals for attacks targeting classifiers at
test time and several other goals can be achieved if the adversary has the
capability of achieving this goal. More specifically, consider a sample $X$ and
a trained DNN resulting in a classifier model $F$. The goal of the adversary is
to produce an adversarial sample $X^*=X+\delta X$ by adding a perturbation
$\delta X$ to sample $X$, such that $F(X^*)=Y^*$ where $Y^*\neq F(X)$ is the
adversarial target output taking the form of an indicator vector for the target
class~\cite{NAS-186}. 

As several approaches at adversarial sample crafting have been proposed in
previous work, we now construct a framework that encompasses these approaches,
for future work to build on. This allows us to compare the strengths and
weaknesses of each method. The resulting crafting framework is illustrated in
Figure~\ref{fig:adversarial-crafting-overview}. Broadly speaking, an adversary
starts by considering a legitimate sample $X$. We assume that the adversary has
the capability of accessing parameters $\theta_F$ of his targeted model $F$ or
of replicating a similar DNN architecture (since adversarial samples are
transferable between DNNs) and therefore has access to its parameter values.
The adversarial sample crafting is then a two-step process: 
\begin{enumerate}
\item Direction Sensitivity Estimation: evaluate the sensitivity of class change to each input feature
\item Perturbation Selection: use the sensitivity information to select a perturbation $\delta X$ among the input dimensions
\end{enumerate}
In other terms, step (1) identifies directions in the data manifold around
sample $X$ in which the model $F$ learned by the DNN is most sensitive and will
likely result in a class change, while step (2) exploits this knowledge to find
an effective adversarial perturbation. Both steps are repeated if necessary, by
replacing $X$ with $X+\delta X$ before starting each new iteration, until the
sample satisfies the adversarial goal: it is classified by deep neural networks
in the target class specified by the adversary using a class indicator vector
$Y^*$.  Note that, as mentioned previously, it is important for the total
perturbation used to craft an adversarial sample from a legitimate sample to be
minimized, at least approximatively. This is essential for adversarial samples
to remain undetected, notably by humans. Crafting adversarial samples using
large perturbations would be trivial. Therefore, if one defines a norm $\|
\cdot \|$ appropriate to describe differences between points in the input
domain of DNN model $F$, adversarial samples can be formalized as a solution to
the following optimization problem:
\begin{equation}
\arg \min_{\delta X} \| \delta X \| \mbox{ s.t. } F(X+\delta X) = Y^*
\end{equation}
Most DNN models $F$ will make this problem non-linear and non-convex, making a
closed-solution hard to find in most cases. We now describe in details our
attack framework approximating the solution to this optimization problem, using
previous work to illustrate each of the two steps.

\textbf{Direction Sensitivity Estimation - } This step considers sample $X$, a
$M$-dimensional input. The goal here is to find the dimensions of $X$ that will
produce the expected adversarial behavior with the smallest perturbation. To
achieve this, the adversary must evaluate the sensitivity of the trained DNN
model $F$ to changes made to input components of $X$. Building such a knowledge
of the network sensitivity can be done in several ways. Goodfellow et
al.~\cite{goodfellow2014explaining} introduced the fast sign gradient method
that computes the gradient of the cost function with respect to the input of
the neural network. Finding sensitivities is then achieved by applying the cost
function to inputs labeled using adversarial target labels. Papernot et
al.~\cite{NAS-186} took a different approach and introduced the forward
derivative, which is the Jacobian of $F$, thus directly providing gradients of
the output components with respect to each input component. Both approaches
define the sensitivity of the network for the given input $X$ in each of its
dimensions~\cite{NAS-186, goodfellow2014explaining}. Miyato et
al.~\cite{miyato2015distributional} introduced another sensitivity estimation
measure, named the \emph{Local Distribution Smoothness}, based on the
Kullback-Leibler divergence, a measure of the difference between two
probability distributions. To compute it, they use an approximation of the
network's Hessian matrix. They however do not present any results on
adversarial sample crafting, but instead focus on using the local distribution
smoothness as a training regularizer improving the classification accuracy. 

\textbf{Perturbation Selection -} The adversary must now use this knowledge
about the network sensitivity to input variations to evaluate which dimensions
are most likely to produce the target misclassification with a minimum total
perturbation vector $\delta X$. Each of the two techniques takes a different
approach again here, depending on the distance metric used to evaluate what a
minimum perturbation is. Goodfellow et al.~\cite{goodfellow2014explaining}
choose to perturb all input dimensions by a small quantity in the direction of
the sign of the gradient they computed. This effectively minimizes the
Euclidian distance between the original and the adversarial samples. Papernot
et al.~\cite{NAS-186} take a different approach and follow a more complex
process involving saliency maps to only select a limited number of input
dimensions to perturb. Saliency maps assign values to combinations of input
dimensions indicating whether they will contribute to the adversarial goal or
not if perturbed. This effectively diminishes the number of input features
perturbed to craft samples. The amplitude of the perturbation added to each
input dimensions is a fixed parameter in both approaches. Depending on the
input nature (images, malware, ...), one method or the other is more suitable
to guarantee the existence of adversarial samples crafted using an acceptable
perturbation $\delta X$. An acceptable perturbation is defined in terms of a
distance metric over the input dimensions (e.g., a $L1, L2$ norm). Depending on
the problem nature, different metrics apply and different perturbation shapes
are acceptable or not.

\subsection{About Neural Network Distillation}

We describe here the approach to distillation introduced by Hinton et
al.~\cite{hinton2015distilling}. Distillation is motivated by the end goal of
reducing the size of DNN architectures or ensembles of DNN architectures, so as
to reduce their computing ressource needs, and in turn allow deployment on
resource constrained devices like smartphones. The general intuition behind the
technique is to extract class probability vectors produced by a first DNN or an
ensemble of DNNs to train a second DNN of reduced dimensionality without loss
of accuracy. 

This intuition is based on the fact that knowledge acquired by DNNs during
training is not only encoded in weight parameters learned by the DNN but is
also encoded in the probability vectors produced by the network. Therefore,
distillation extracts class knowledge from these probability vectors to
transfer it into a different DNN architecture during training. To perform this
transfer, distillation labels inputs in the training dataset of the second DNN
using their classification predictions according to the first DNN.  The benefit
of using class probabilities instead of hard labels is intuitive as
probabilities encode additional information about each class, in addition to
simply providing a sample's correct class. Relative information about classes
can be deduced from this extra entropy. 

To perform distillation, a large network whose output layer is a softmax is
first trained on the original dataset as would usually be done. An example of
such a network is depicted in Figure~\ref{fig:dnn-architecture}. A softmax
layer is merely a layer that considers a vector $Z(X)$ of outputs produced by
the last hidden layer of a DNN, which are named \emph{logits}, and normalizes
them into a probability vector $F(X)$, the ouput of the DNN, assigning a
probability to each class of the dataset for input $X$. Within the softmax
layer, a given neuron corresponding to a class indexed by $i\in 0..N-1$ (where
$N$ is the number of classes) computes component $i$ of the following output
vector $F(X)$: 
\begin{equation}
\label{eq:softmax}
 F(X) = \left[ \frac{e^{z_i(X)/T}}{\sum_{l=0}^{N-1} e^{z_l(X)/T}} \right]_{i\in0 .. N-1}
\end{equation}
where $Z(X)=z_0(X), ..., z_{N-1}(X)$ are the $N$ logits corresponding to the
hidden layer outputs for each of the $N$ classes in the dataset, and $T$ is a
parameter named \emph{temperature} and shared across the softmax layer.
Temperature plays a central role in underlying phenomena of distillation as we
show later in this section. In the context of distillation, we refer to this
temperature as the \emph{distillation temperature}. The only constraint put on
the training of this first DNN is that a high temperature, larger than 1,
should be used in the softmax layer.

The high temperature forces the DNN to produce probability vectors with
relatively large values for each class. Indeed, at high temperatures, logits in
vector $Z(X)$ become negligible compared to temperature $T$. Therefore, all
components of probability vector $F(X)$ expressed in Equation~\ref{eq:softmax}
converge to $1/N$ as $T\rightarrow \infty$. The higher the temperature of a
softmax is, the more ambiguous its probability distribution will be (i.e. all
probabilities of the output $F(X)$ are close to $1/N$), whereas the smaller the
temperature of a softmax is, the more discrete its probability distribution
will be (i.e. only one probability in output $F(X)$ is close to $1$ and the
remainder are close to $0$). 

The probability vectors produced by the first DNN are then used to label the
dataset. These new labels are called \emph{soft labels} as opposed to
\emph{hard class labels}. A second network with less units is then trained
using this newly labelled dataset. Alternatively, the second network can also
be trained using a combination of the hard class labels and the probability
vector labels. This allows the network to benefit from both labels to converge
towards an optimal solution. Again, the second network is trained at a high
softmax temperature identical to the one used in the first network. This second
model, although of smaller size, achieves comparable accuracy than the original
model but is less computationally expensive. The  temperature is set back to 1
at test time so as to produce more discrete probability vectors during
classification. 


\section{Defending DNNs using Distillation}
\label{sec:defense}

Armed with background on DNNs in adversarial settings, we now introduce a
defensive mechanism to reduce vulnerabilities exposing DNNs to adversarial
samples. We note that most previous work on combating adversarial samples
proposed regularizations or dataset augmentations. We instead take a radically
different approach and use {\em distillation}, a training technique described in the previous section, to improve the robustness of DNNs. We describe how we adapt distillation into
\emph{defensive distillation} to address the problem of DNN vulnerability to
adversarial perturbations. We provide a justification of the approach using
elements from learning theory.

\subsection{Defending against Adversarial Perturbations}

To formalize our discussion of defenses against adversarial samples, we now
propose a metric to evaluate the resilience of DNNs to adversarial noise. To
build an intuition for this metric, namely the \emph{robustness} of a network,
we briefly comment on the underlying vulnerabilities exploited by the attack
framework presented above. We then formulate requirements for defenses capable
of enhancing  classification robustness.

In the framework discussed previously, we underlined the fact that attacks
based on adversarial samples were primarily exploiting gradients computed to
estimate the sensitivity of networks to its input dimensions. To simplify our
discussion, we refer to these gradients as \emph{adversarial gradients} in the
remainder of this document.  If adversarial gradients are high, crafting
adversarial samples becomes easier because small perturbations will induce high
network output variations. To defend against such perturbations, one must
therefore reduce these variations around the input, and consequently the
amplitude of adversarial gradients. In other words, we must smooth the model
learned during training by helping the network generalize better to samples
outside of its training dataset. Note that adversarial samples are not
necessarily found in ``nature", because adversarial samples are specifically
crafted to break the classification learned by the network. Therefore, they are
not necessarily extracted from the input distribution that the DNN architecture
tries to model during training. 

\begin{figure}
\centering
\includegraphics[width=3in]{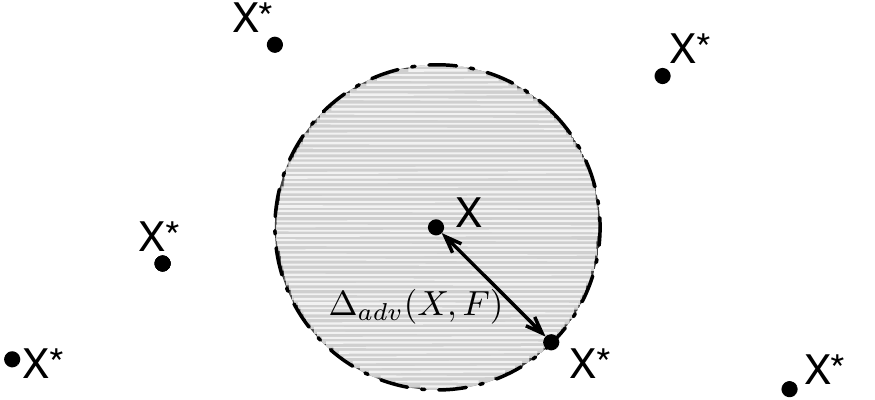}
\caption{\textbf{Visualizing the hardness metric:} This 2D representation illustrates the hardness metric as the radius of the disc centered at the original sample $X$ and going through the closest adversarial sample $X^*$ among all the possible adversarial samples crafted from sample $X$. Inside the disc, the class output by the classifier is constant. However, outside the disc, all samples $X^*$ are classified differently than $X$.}
\label{fig:robustness}
\end{figure}

\textbf{DNN Robustness -} We informally defined the notion of \emph{robustness}
of a DNN to adversarial perturbations as its capability to resist
perturbations. In other words, a robust DNN should (i) display good accuracy inside
and outside of its training dataset as well as (ii) model a smooth classifier
function $F$ which would intuitively classify inputs relatively consistently in
the neighborhood of a given sample. The notion of neighborhood can be defined
by a norm appropriate for the input domain. Previous work has formalized a
close definition of robustness in the context of other machine learning
techniques~\cite{fawzi2015analysis}.  The intuition behind this metric is that
robustness is achieved by ensuring that the classification output by a DNN
remains somewhat constant in a closed neighborhood around any given sample
extracted from the classifier's input distribution. This idea is illustrated in
Figure~\ref{fig:robustness}. The larger this neighborhood is for all inputs
within the natural distribution of samples, the more robust is the DNN. Not all
inputs are considered, otherwise the ideal robust classifier would be a
constant function, which has the merit of being very robust to adversarial
perturbations but is not a very interesting classifier. We extend the
definition of robustness introduced in~\cite{fawzi2015analysis} to the
adversarial behavior of source-target class pair misclassification within the
context of classifiers built using DNNs. The robustness of a trained DNN model
$F$ is: 
\begin{equation}
\label{eq:robustness-metric}
\rho_{adv}(F)=E_{\mu}[\Delta_{adv}(X,F)]
\end{equation} 
where inputs $X$ are drawn from distribution $\mu$ that DNN architecture is attempting to model with $F$, and $\Delta_{adv}(X,F)$ is defined to be the minimum perturbation required to misclassify sample $x$ in each of the other classes:
\begin{equation}
\label{eq:min-misclassification-perturbation}
\Delta_{adv}(X,F)=\arg \min_{\delta X} \{ \|\delta X\| : F(X+\delta X) \neq  F(X) \}
\end{equation} 
where $\|\cdot\|$ is a norm and must be specified accordingly to the context. The higher the average minimum perturbation required to misclassify a sample from the data manifold is, the more robust the DNN is to adversarial samples. 

\textbf{Defense Requirements - }Pulling from this formalization of DNN robustness. we now outline design requirements for defenses against adversarial perturbations:
\begin{itemize}
\item \emph{Low impact on the architecture:} techniques introducing limited, 
 modifications to the architecture 
are preferred in our approach because introducing new architectures not studied 
in the literature requires analysis of their behaviors. Designing new architectures
and benchmarking them against our approach is left as future work. 
\item \emph{Maintain accuracy}: defenses against adversarial samples should not decrease the DNN's classification accuracy. This discards solutions based on weight decay, through $L1, L2$ regularization, as they will cause underfitting.
\item \emph{Maintain speed of network:} the solutions should not significantly
impact the running time of the classifier at test time. Indeed, running time at
test time matters for the usability of DNNs, whereas an impact on training time
is somewhat more acceptable because it can be viewed as a fixed cost. Impact on
training should nevertheless remain limited to ensure DNNs can still take
advantage of large training datasets to achieve good accuracies. For instance,
solutions based on Jacobian regularization, like double
backpropagation~\cite{drucker1992improving}, or using radial based activation
functions~\cite{goodfellow2014explaining} degrade DNN training performance. 
\item \emph{Defenses should work for adversarial samples relatively close to
points in the training dataset}~\cite{goodfellow2014explaining, NAS-186}.
Indeed, samples that are very far away from the training dataset, like those
produced in~\cite{nguyen2014deep}, are irrelevant to security because they can
easily be detected, at least by humans. However, limiting sensitivity to
infinitesimal perturbation (e.g., using double
backpropagation~\cite{drucker1992improving}) only provides constraints very
near training examples, so it does not solve the adversarial perturbation
problem. It is also very hard or expensive to make derivatives smaller to limit
sensitivity to infinitesimal perturbations. 
\end{itemize}

We show in our approach description below that our defense technique does not
require any modification of the neural network architecture and that it has a
low overhead on training and no overhead on test time. In the evaluation
conducted in section~\ref{sec:evaluation}, we also show that our defense
technique fits the remaining defense requirements by evaluating the accuracy of
DNNs with and without our defense deployed, and studying the generalization
capabilities of networks to show how the defense impacted adversarial samples.

\begin{figure*}
\centering
\includegraphics[width=\textwidth]{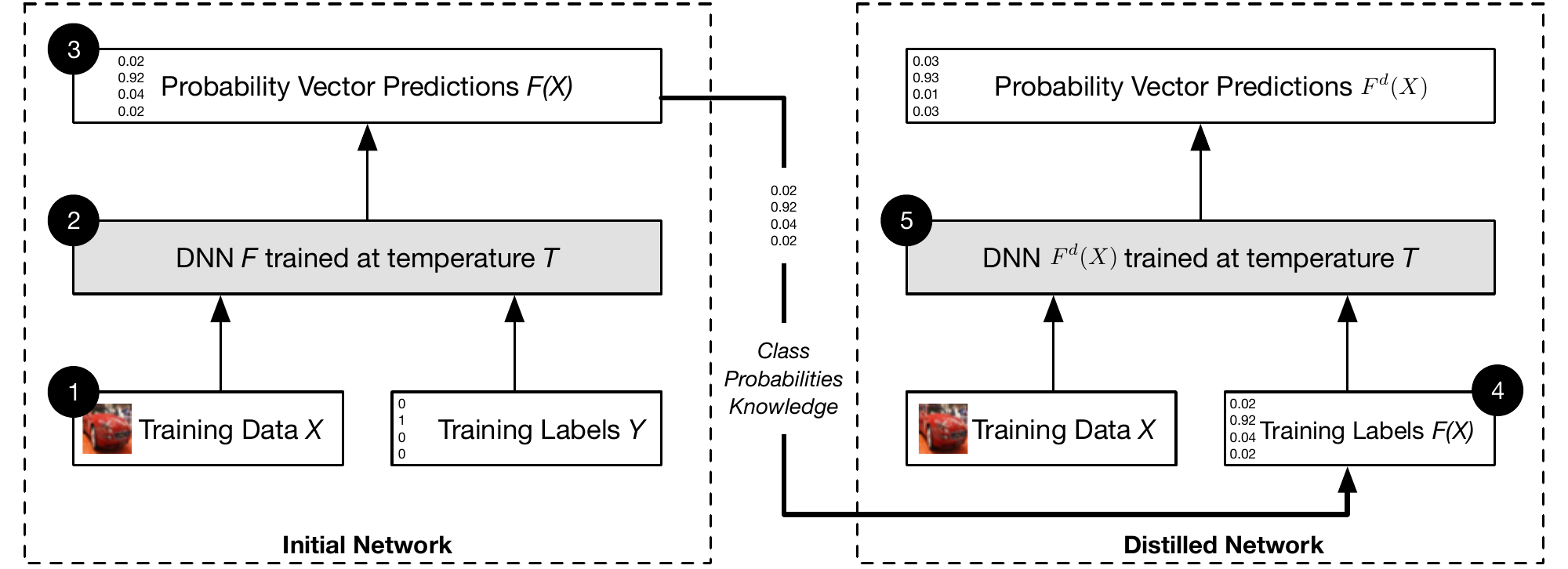}
\caption{\textbf{An overview of our defense mechanism based on a transfer of
knowledge contained in probability vectors through distillation:} We first
train an initial network $F$ on data $X$ with a softmax temperature of $T$. We
then use the probability vector $F(X)$, which includes additional knowledge
about classes compared to a class label, predicted by network $F$ to train a
distilled network $F^d$ at temperature $T$ on the same data $X$.}
\label{fig:distillation-overview}
\end{figure*}

\subsection{Distillation as a Defense}

We now introduce \emph{defensive distillation}, which is the technique we
propose as a defense for DNNs used in adversarial settings, when adversarial
samples cannot be permitted.  Defensive distillation is adapted from the
distillation procedure, presented in
section~\ref{sec:adversarial-deep-learning}, to suit our goal of improving DNN
classification resilience in the face of adversarial perturbations.

Our intuition is that knowledge extracted by distillation, in the form of
probability vectors, and  transferred in smaller networks to maintain
accuracies comparable with those of larger networks can also be beneficial to
improving  generalization capabilities of DNNs outside of their training
dataset and therefore enhances their resilience to perturbations. Note that
throughout the remainder of this paper, we assume that considered DNNs are used
for classification tasks and designed with a softmax layer as their output
layer. 

The main difference between defensive distillation and the original
distillation proposed by Hinton et al.~\cite{hinton2015distilling} is that we
keep the same network architecture to train both the original network as well
as the distilled network. This difference is justified by our end which is
resilience instead of compression. The resulting defensive distillation
training procedure is illustrated in Figure~\ref{fig:distillation-overview} and
outlined as follows:
\begin{enumerate}
\item The input of the defensive distillation training algorithm is a set $\cal X$ of samples with their class labels. Specifically, let $X \in {\cal X}$ be a sample,
  we use $Y(X)$ to denote its discrete label, also referred to as \emph{hard label}. $Y(X)$ is an indicator vector such that the only non-zero element corresponds to the correct class' index (e.g. $(0, 0, 1, 0, \dots, 0)$ indicates that the sample is in the class with index $2$).
\item Given this training set $\{ (X,Y(X)):X\in \cal{X} \}$, we train a deep neural network $F$
  with a softmax output layer at temperature $T$.
  As we discussed before, $F(X)$ is a probability vector over the class of
  all possible labels. More precisely, if the model $F$ has parameters $\theta_F$,
  then its output on $X$ is a probability distribution $F(X)=p(\cdot|X,\theta_F)$,
  where for any label $Y$ in the label class,
  $p(Y|X,\theta_F)$ gives a probability that the label is $Y$.
  To simplify our notation later, we use $F_i(X)$ to denote the probability of input $X$ to be in class $i\in 0..N-1$ according to model $F$ with parameters $\theta_F$.
  \item We form a new training set, by consider samples of the form $(X,F(X))$
  for $X \in {\cal X}$. That is, instead of using hard class label $Y(X)$ for $X$,
  we use the \emph{soft-target} $F(X)$ encoding $F$'s belief probabilities
  over the label class.
\item Using the new training set $\{(X, F(X)) : X \in {\cal X}\}$
  we then train another DNN model $F^d$,
  with the same neural network architecture as $F$,
  and the temperature of the softmax layer remains $T$.
  This new model is denoted as $F^d$ and referred to as the \emph{distilled model}.
\end{enumerate}

Again, the benefit of using soft-targets $F(X)$ as training labels lies in the
additional knowledge found in probability vectors compared to hard class
labels.  This additional entropy encodes the relative differences between
classes. For instance, in the context of digit recognition developed later in
section~\ref{sec:evaluation}, given an image $X$ of some handwritten digit,
model $F$ may evaluate the probability of class $7$ to  $F_7(X)=0.6$  and the
probability of label $1$ to $F_1(X)=0.4$, which then indicates some structural
similarity between $7$s and $1$s.

Training a network with this explicit relative information about classes
prevents models from fitting too tightly to the data, and contributes to a
better generalization around training points. Note that the knowledge
extraction performed by distillation is controlled by a parameter: the softmax
temperature $T$. As described in section~\ref{sec:adversarial-deep-learning},
high temperatures force DNNs to produce probabilities vectors with large values
for each class. In sections~\ref{sec:analysis} and~\ref{sec:evaluation}, we
make this intuition more precise with a theoretical analysis and an empirical
evaluation.


\section{Analysis of Defensive Distillation}
\label{sec:analysis}

We now explore analytically the impact of defensive distillation on DNN
training and resilience to adversarial samples. As stated above, our intuition
is that probability vectors produced by model $F$ encode supplementary entropy
about classes that is beneficial during the training of distilled model $F^d$.
Before proceeding further, note that our purpose in this section is not to
provide a definitive argument about using defensive distillation to combat
adversarial perturbations, but rather we view it as an initial step towards
drawing a connection between distillation, learning theory, and DNN robustness
for future work to build upon. This analysis of distillation is split in three
folds studying (1) network training, (2) model sensitivity, and (3)
the generalization capabilities of a DNN.

Note that with training, we are looking to converge towards a function $F^*$
resilient to adversarial noise and capable of generalizing better. The
existence of function $F^*$ is guaranteed by the universality theorem for
neural networks~\cite{Cybenko92}, which states that with enough neurons and
enough training points, one can approximate any continuous function with
arbitrary precision. In other words, according to this theorem, we know that
there exists a neural network architecture that converges to $F^*$ if it is
trained on a sufficient number of samples. With this result in mind, a natural
hypothesis is that distillation helps convergence of DNN models towards the
optimal function $F^*$ instead of a different local optimum during training.

\subsection{Impact of Distillation on Network Training}
To precisely understand the effect of defensive distillation on adversarial
crafting, we need to analyze more in depth the training process. Throughout
this analysis, we frequently refer to the training steps for defensive
distillation, as described in Section~\ref{sec:defense}. Let us start by
considering the training procedure of the first model $F$, which corresponds to
step (2) of defensive distillation. Given a batch of samples $\{ (X, Y(X))\ |\
X \in {\cal X} \}$ labeled with their correct classes, training algorithms typically aim
to solve the following optimization problem:
\begin{equation}
\label{eq:loglike}
\arg\min_{\theta_F}
- \frac{1}{|{\cal X}|}\sum_{X \in {\cal X}}\sum_{i\in 0..N} Y_i(X)\log{F_i(X)}.
\end{equation}
where $\theta_F$ is the set of parameters of model $F$ and $Y_i$ is the $i^{th}$ component of $Y$. That is,
for each sample $(X, Y(X))$ and hypothesis, i.e. a model $F$ with parameters
$\theta_F$, we consider the log-likelihood $\ell(F, X, Y(X)) = -Y(X)\cdot \log{ F(X)}$ of $F$ on $(X, Y(X))$ and average it over the
entire training set $\cal{X}$. Very roughly speaking, the goal of this
optimization is to
adjust the weights of the model so as to push each $F(X)$ towards $Y(X)$.
However, readers will notice that since $Y(X)$ is an indicator vector of input
$X$'s class, Equation~\ref{eq:loglike} can be simplified to: 
\begin{equation}
\label{eq:loglike-2}
\arg\min_{\theta_F}
- \frac{1}{|{\cal X}|}\sum_{X \in {\cal X}}\log{F_{t(X)}(X)}.
\end{equation}
where $t(X)$ is the only element in indicator vector $Y(X)$ that is equal to
$1$, in other words the index of the sample's class. This means that when
performing updates to $\theta_F$, the training algorithm will constrain any
output neuron different from the one corresponding to probability $F_{t(X)}(X)$ to give a $0$ output. 
However, this forces the DNN to make overly confident predictions in the 
sample class. We
argue that this is a fundamental lack of precision during training as most of
the architecture remains unconstrained as weights are updated.

Let us move on to explain how defensive distillation solves this issue, and how
the distilled model $F^d$ is trained. As mentioned before, while the original
training dataset is $\{ (X, Y(X)) : X \in {\cal X} \}$, the distilled model
$F^d$ is trained using the same set of samples but labeled with soft-targets
$\{(X, F(X)) : X \in {\cal X}\}$ instead. This set is constructed at step (3)
of defensive distillation. In other words, the label of $X$ is no longer the
indicator vector $Y(X)$ corresponding to the hard class label of $X$, but
rather the soft label of input $X$: a probability vector $F(X)$. Therefore,
$F^d$ is trained, at step (4), by solving the following optimization problem:
\begin{equation}
\arg\min_{\theta_F} - 
\frac{1}{|{\cal X}|}
\sum_{X \in {\cal X}}
\sum_{i\in0..N} F_i(X)\log{F_i^d(X)} 
\end{equation}
Note that the key difference here is that because we are using soft labels
$F(X)$, it is not trivial anymore that most components of the double sum are
null. Instead, using probabilities $F_j(X)$ ensures that the training algorithm
will constrain all output neurons $F_j^d(X)$ proportionally to their 
likelihood when updating parameters
$\theta_F$. We argue that this contributes to improving the generalizability of
classifier model $F$ outside of its training dataset, by avoiding situations
where the model is forced to make an overly confident prediction in one class
when a sample includes characteristics of two or more classes (for instance,
 when classifying digits, an instance of a 8 include shapes also characteristic of a digit 3). 

Note that model $F^d$ should theoretically eventually
converge towards $F$. Indeed, locally at each
point $(X, F(X))$, the optimal solution is for model $F^d$ to be such that
$F^d(X) = F(X)$. To see this, we observe that training aims to minimize the
cross entropy between $F^d(X)$ and $F(X)$, which is equal to:
\begin{equation}
{\rm H}\left(F^d(X), F(X)\right) = 
 {\rm H}(F(X))
 + {\rm KL}\big( F(X)\ \|\ F^d(X) \big)
\end{equation}
where ${\rm H}(F(X))$ is the Shannon entropy of
$F(X)$) and ${\rm KL}$ denotes the Kullback-Leibler divergence. Note that this
quantity is minimized when the KL divergence is equal to $0$, which is only
true when $F^d(X) = F(X)$. Therefore, an ideal training procedure would result
in model $F^d$ converging to the first model $F$. However, empirically this is
not the case because training algorithms approximate the solution of the
training optimization problem, which is often non-linear and non-convex.
Furthermore, training algorithms only have access to a finite number of
samples. Thus, we do observe empirically  a better behavior in adversarial
settings from model $F^d$ than model $F$. We confirm this result in
Section~\ref{sec:evaluation}.

\subsection{Impact of Distillation on Model Sensitivity} 
Having studied the impact of defensive distillation on optimization problems solved during DNN
training, we now further investigate why adversarial perturbations are harder to craft
on DNNs trained with defensive distillation at high temperature. The goal of our analysis here is to
provide an intuition of how distillation at high temperatures improves the smoothness of the distilled 
model $F^d$ compared to model $F$, thus reducing its sensitivity to small input variations.

The model's sensitivity to input variation is quantified by its Jacobian.
We first show why the amplitude of Jacobian's components naturally decrease as 
the temperature of the softmax increases. Let us derive the expression of component $(i,j)$ of the Jacobian for a model $F$ at temperature $T$:
\begin{equation}
\left.\frac{\partial F_i(X)}{\partial X_j}\right|_T
= \frac{\partial}{\partial X_j}\left(
\frac{e^{z_i(X)/T}}{\sum_{l=0}^{N-1} e^{z_l(X)/T}}\right)
\end{equation}
where $z_0(X), \dots, z_{N-1}(X)$ are the inputs to the softmax layer---also referred to as logits---and are simply the outputs of the last hidden layer of model $F$. For the sake of notation clarity, we do not write the dependency of $z_0(X), \dots, z_{N-1}(X)$ to $X$ and simply write $z_0, \dots, z_{N-1}$. Let us also write $g(X) = \sum_{l=0}^{N-1} e^{z_l(X)/T}$, we then have:
\begin{align*}
 \left. \frac{\partial F_i(X)}{\partial X_j}\right|_T = &\frac{\partial}{\partial X_j}\left(
\frac{e^{z_i/T}}{\sum_{l=0}^{N-1} e^{z_l/T}}\right) \\
= & \frac{1}{g^2(X)}
\left(
  \frac{\partial e^{z_i(X)/T}}{\partial X_j}g(X)
  - e^{z_i(X)/T}\frac{\partial g(X)}{\partial X_j}
\right)  \\
= & \frac{1}{g^2(X)} \frac{e^{z_i/T}}{T}
\left(
 \sum_{l=0}^{N-1}\frac{\partial z_i}{\partial X_j}e^{z_l/T}
  -\sum_{l=0}^{N-1}\frac{\partial z_l}{\partial X_j}e^{z_l/T}
\right) \\
= & \frac{1}{T} \frac{e^{z_i/T}}{g^2(X)}
\left(
  \sum_{l=0}^{N-1}
  \left(
    \frac{\partial z_i}{\partial X_j}-\frac{\partial z_l}{\partial X_j}
  \right)e^{z_l/T}
\right)
\end{align*}
The last equation yields that increasing the softmax temperature
$T$ for fixed values of the logits $z_0, \dots, z_{N-1}$ will reduce the absolute value of all components of model $F$'s Jacobian
matrix because (i) these components are inversely proportional to
temperature $T$, and (ii) logits are divided by temperature $T$ before being
exponentiated. 

This simple analysis shows how using high temperature systematically reduces the 
model sensitivity to small variations of its inputs 
when defensive distillation is performed at training time. However, at test time, the temperature is 
decreased back to $T=1$ in order to make predictions on unseen inputs. Our intuition
is that this does not affect the model's sensitivity as weights learned
during training will not be modified by this change of temperature, and 
decreasing temperature only makes the class probability vector more discrete,
 without changing the relative ordering of classes. In a 
way, the smaller sensitivity imposed by using a high temperature is 
encoded in the weights during training and is thus still observed at test time. 
While this explanation matches both our intuition and the experiments detailed
later in section~\ref{sec:evaluation}, further formal analysis is needed.  We plan to pursue 
this in future work.

\subsection{Distillation and the Generalization Capabilities of DNNs} 

We now provide elements of learning theory to
analytically understand the impact of distillation on generalization
capabilities. We formalize our intuition that models 
benefit from soft labels. Our motivation stems from the fact that not only do
probability vectors $F(X)$ encode model $F$'s knowledge regarding the correct
class of $X$, but it also encodes the knowledge of how classes are likely,
relatively to each other.

Recall our example of handwritten digit recognition. Suppose we are given a
sample  $X$ of some hand-written $7$ but that the writing is so bad that the
$7$ looks like a $1$. Assume a model $F$ assigns probability $F_7(X)=0.6$ on
$7$ and probability $F_1(X)= 0.4$ on $1$, when given sample $X$ as an input.
This indicates that $7$s and $1$s look similar and intuitively allows a model
to learn the \emph{structural similarity} between the two digits. In contrast,
a hard label leads the model to believe that $X$ is a $7$, which can be
misleading since the sample is poorly written. This example illustrate the need
for algorithms not fitting too tightly to particular samples of $7$s, which in
turn prevent models from overfitting and offer better generalizations.

To make this intuition more precise, we resort to the recent breakthrough in
computational learning theory on the connection between learnability and
stability.  Let us first present some elements of stable learning theory
to facilitate our discussion. Shalev-Schwartz et al.~\cite{SSSSS10} proved that
learnability is equivalent to the existence of a learning rule that is
simultaneously an asymptotic empirical risk minimizer and stable. More
precisely, let $(Z = X \times Y, {\cal H}, \ell)$ be a learning problem where
$X$ is the input space, $Y$ is the output space, $\cal H$ is the hypothesis
space, and $\ell$ is an instance loss function that maps a pair $(w, z) \in
{\cal H} \times Z$ to a positive real loss. Given a training set $S = \{z_i : i
\in [n]\}$, we define the empirical loss of a hypothesis $w$ as $L_S(w) =
\frac{1}{n}\sum_{i \in [n]}\ell(w, z_i)$. We denote the minimal empirical risk
as $L_S^* = \min_{w \in {\cal H}}L_S(w)$. We are ready to present the following
two definitions:
\begin{definition}[Asymptotic Empirical Risk Minimizer]
  A learning rule $A$ is an {\em asymptotic empirical risk minimizer},
   if there is a rate function\footnote{A function that non-increasingly vanishes to $0$ as $n$ grows.} $\varepsilon(n)$
  such that
  for every training set $S$ of size $n$,
  \[ L_S(A(S)) - L_S^* \le \varepsilon(n). \]
\end{definition}

\begin{definition}[Stability]
  We say that a learning rule $A$ is $\varepsilon(n)$ stable   if for every two training sets $S,S'$ that only differ in one training item,   and for every $z \in Z$,  
   \[ |\ell(A(S), z) - \ell(A(S'), z)| \le \varepsilon(n) \]
  where $h = A(S)$ is the output of $A$ on training set $S$,  and $\ell(A(S), z) = \ell(h, z)$ denotes the loss of $h$ on $z$.
\end{definition}
An interesting result to progress in our discussion is the following Theorem mentioned previously and proved in~\cite{SSSSS10}.
\begin{theorem}
  \label{thm:aerm-and-stable-implies-learnable}
  If there is a learning rule $A$ that is both an asymptotic empirical risk minimizer and stable,
  then $A$ generalizes, which means that   the generalization error
  $L_{\cal D}(A(S))$ converges to $L_{\cal D}^*
  = \min_{h \in {\cal H}}L_{\cal D}(h)$
  with some rate $\varepsilon(n)$ independent of
  any data generating distribution $\cal D$.
\end{theorem}
We now link this theorem back to our discussion. We observe that, by
appropriately setting the temperature $T$, it follows that for any datasets $S, S'$ only differing by
one training item, the new generated training sets $(X, F^S(X))$ and $(X,
F^{S'}(X))$ satisfy a very strong stability condition. This in turn means that
for any $X \in {\cal X}$, $F^S(X)$ and $F^{S'}(X)$ are statistically close.
Using this observation, one can note that defensive distillation training
satisfies the stability condition defined above. 
 
Moreover, we
deduce from the objective function of defensive distillation that the approach
minimizes the empirical risk. Combining these two results together with
Theorem~\ref{thm:aerm-and-stable-implies-learnable} allows us to conclude that
the distilled model generalizes well.

We conclude this discussion by noting that we did not strictly prove that the distilled
model generalizes better than a model trained without defensive distillation.
This is right and indeed this property is difficult to prove when dealing with
DNNs because of the non-convexity properties of optimization problems solved
during training. To deal with this lack of convexity, approximations are made
to train DNN architectures and model optimality cannot be guaranteed. To the
best of our knowledge, it is difficult to argue the learnability of DNNs in the
first place, and no good learnability results are known. However, we do believe
that our argument provides the readers with an intuition of why distillation
may help generalization. 


\section{Evaluation}
\label{sec:evaluation}


\begin {table}
\centering
\begin{tabular}{|p{3.0cm} |p{2.1cm}| p{2.3cm}|}
\hline
\multirow{2}{3cm}{\textbf{Layer Type}} & \multirow{2}{2.1cm}{\textbf{MNIST\\ Architecture}}  & \multirow{2}{2.3cm}{\textbf{CIFAR10 Architecture}}   \\ 
 &   &   \\ \hline \hline
Relu Convolutional  &  32 filters (3x3) & 64 filters (3x3)  \\ \hline 
Relu Convolutional  &  32 filters (3x3) & 64 filters (3x3)  \\ \hline 
Max Pooling  &  2x2 & 2x2  \\ \hline 
Relu Convolutional  &  64 filters (3x3) & 128 filters (3x3)  \\ \hline 
Relu Convolutional  &  64 filters (3x3) & 128 filters (3x3)  \\ \hline 
Max Pooling  &  2x2 & 2x2  \\ \hline 
Relu Fully Connect.  & 200 units & 256 units \\ \hline 
Relu Fully Connect.  & 200 units & 256 units\\ \hline 
Softmax  & 10 units &  10 units \\ \hline 
\end{tabular}
\caption{\textbf{Overview of Architectures:} both architectures are based on a
succession of 9 layers. However, the MNIST architecture uses less units in each
layers than the CIFAR10 architecture because its input is composed of less
features. }
\label{tbl:eval-architectures}
\end{table}

\begin{table}
\centering
\begin{tabular}{|p{3cm} |p{2.1cm}| p{2.3cm}|}
\hline
\multirow{2}{2.3cm}{\textbf{Parameter}} & \multirow{2}{2.3cm}{\textbf{MNIST Architecture}}  & \multirow{2}{2.3cm}{\textbf{CIFAR10 Architecture}}  \\  
 &   &   \\ \hline \hline
Learning Rate & 0.1  & 0.01 (decay 0.5) \\ \hline 
Momentum & 0.5  & 0.9 (decay 0.5)\\ \hline
Decay Delay & -  & 10 epochs \\ \hline 
Dropout Rate (Fully Connected Layers) & 0.5  & 0.5 \\ \hline 
Batch Size & 128 & 128 \\ \hline 
Epochs & 50  & 50 \\ \hline 
\end{tabular}
\caption{\textbf{Overview of Training Parameters:} the CIFAR10 architecture
training was slower than the MNIST architecture and uses parameter decay to
ensure model convergence. }
\label{tbl:training-hyper}
\end{table}

This section empirically evaluates defensive distillation, using two DNN network architectures.  The central asked questions and results of this emprical study include:

\begin{itemize}
\itembase{3pt}

\item {\it Q: Does defensive distillation improve resilience against
adversarial samples while retaining classification accuracy?} (see
Section~\ref{sec:defdistill}) -
{\bf Result}: Distillation reduces the success rate of adversarial crafting
from $95.89\%$ to $0.45\%$ on our first DNN and dataset, and from $87.89\%$ to
$5.11\%$ on a second DNN and dataset. Distillation has negligible or non
existent degradation in model classification accuracy in these settings. Indeed
the accuracy variability between models trained without distillation and with
distillation is smaller than $1.37\%$ for both DNNs.  

\item 
{\it Q: Does defensive distillation reduce DNN sensitivity to inputs?} (see
Section~\ref{sec:sensitivity})
{\bf Result}: Defensive distillation reduces DNN sensitivity to input
perturbations, where experiments show that performing distillation at high
temperatures can lead to decreases in the amplitude of adversarial gradients by
factors up to $10^{30}$.

\item 
{\it Q: Does defensive distillation lead to more robust DNNs?} (see
Section\ref{sec:robust})
{\bf Result}: Defensive distillation impacts the average minimum percentage of
input features to be perturbed to achieve adversarial targets (i.e.,
robustness). In our DNNs, distillation increases robustness by $790\%$ for the
first DNN and $556\%$ for the second DNN: on our first network the metric
increases from $1.55\%$ to $14.08\%$ of the input features, in the second
network the metric increases from $0.39\%$ to $2.57\%$.

\end{itemize}

\vspace*{-0.15in}

\subsection{Overview of the Experimental Setup}

\begin{figure}
\centering
\includegraphics[width=3.2in]{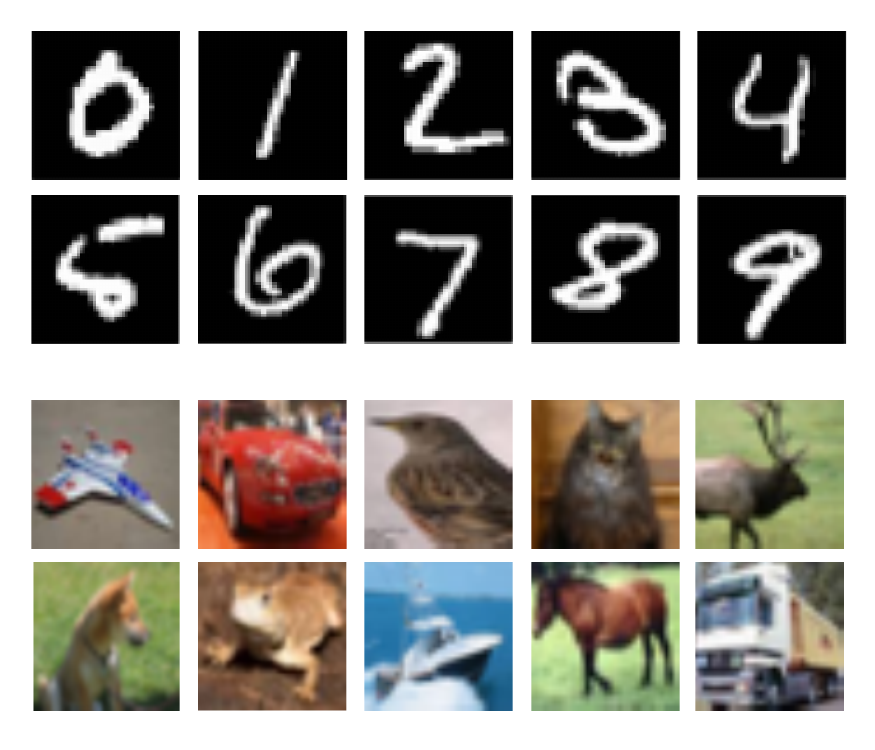}
\caption{\textbf{Set of legitimate samples:} these samples were extracted from
each of the 10 classes of the MNIST handwritten digit dataset (top) and CIFAR10
image dataset (bottom).}
\label{fig:samples}
\end{figure}

\textbf{Dataset Description -} All of the experiments described in this section
are performed on two canonical machine learning datasets: the
MNIST~\cite{lecun1998mnist} and CIFAR10~\cite{krizhevsky2009learning} datasets.
The MNIST dataset is a collection of $70,000$ black and white images of
handwritten digits, where each pixel is encoded as a real number between $0$ and $1$. 
The samples are split between a training set of $60,000$
samples and a test set of $10,000$. The classification goal is to determine the
digit written. The classes therefore range from 0 to 9. The CIFAR10 dataset is
a collection of $60,000$ color images. Each pixel is encoded by 3 color 
components, which after preprocessing have values in $[-2.22,2.62]$ for the 
test set. The samples are split between a training
set of $50,000$ samples and a test set of $10,000$ samples. The images are to
be classified in one of the 10 mutually exclusive classes: airplane,
automobile, bird, cat, deer, dog, frog, horse, ship, and truck. Some
representative samples from each dataset are shown in Figure~\ref{fig:samples}. 

\textbf{Architecture Characteristics -} We implement two deep neural network
architectures whose specificities are described in
Table~\ref{tbl:eval-architectures} and training hyper-parameters included in
Table~\ref{tbl:training-hyper}: the first architecture is a 9 layer
architecture trained on the MNIST dataset, and the second architecture  is a 9
layer architecture trained on the CIFAR10 dataset. The architectures are based
on convolutional neural networks, which have been widely studied in the
literature. We use momentum and parameter decay to ensure model convergence, and dropout to prevent overfitting. Our DNN performance is consistent with
DNNs that have evaluated these datasets before. 

The MNIST architecture is constructed using 2 convolutional layers with 32
filters followed by a max pooling layer, 2 convolutional layers with 64 filters
followed by a max pooling layer, 2 fully connected layers with 200 rectified
linear units, and a softmax layer for classification in 10 classes. The
experimental DNN is trained on batches of $128$ samples with a learning rate of
$\eta=0.1$ for $50$ epochs.  The resulting DNN achieves a $99.51\%$ correct
classification rate on the data set, which is comparable to state-of-the-art
DNN accuracy. 

The CIFAR10 architecture is a succession of 2 convolutional layers with 64
filters followed by a max pooling layer, 2 convolutional layers with 128
filters followed by a max pooling layer, 2 fully connected layers with 256
rectified linear units, and a softmax layer for classification. When trained on
batches of $128$ samples for the CIFAR10 dataset with a learning rate of
$\eta=0.01$ (decay of $0.95$ every $10$ epochs), a momentum of $0.9$ (decay of
$0.5$ every $10$ epochs) for $50$ epochs, a dropout rate of $0.5$, the
architecture achieves a $80.95\%$ accuracy on the CIFAR10 test set, which is
comparable to state-of-the-art performance for unaugmented datasets. 

To train and use DNNs, we use Theano~\cite{bergstra2010theano}, which is
designed to simplify large-scale scientific computing, and
Lasagne~\cite{lasagne}, which simplifies the design and implementation of deep
neural networks using computing capabilities offered by Theano. This setup
allows us to efficiently implement network training as well as the computation
of gradients needed to craft adversarial samples. We configure Theano to make
computations with float32 precision, because they can then be accelerated using
graphics processing. Indeed, we use machines equipped with Nvidia
Tesla K5200 GPUs. 

\vspace*{0.05in}

\textbf{Adversarial Crafting -} We implement adversarial sample crafting as
detailed in~\cite{NAS-186}. The adversarial goal is to alter any sample $X$
originally classified in a source class $F(X)$ by DNN $F$ so as to have the
perturbed sample $X^*$ classified by DNN $F$ in a distinct target class $F(X^*)
\neq F(X)$. To achieve this goal, the attacker first computes the Jacobian of
the neural network output with respect to its input. A perturbation is then
constructed by ranking input features to be perturbed using a saliency map
based on the previously computed network Jacobian and giving preference to
features more likely to alter the network output. Each feature perturbed is set to
$1$ for the MNIST architecture and $2$ for the CIFAR10 dataset. Note that the 
attack~\cite{NAS-186} we implemented in this evaluation is based on 
perturbing very few pixels by a large amount, while previous 
attacks~\cite{szegedy2013intriguing, goodfellow2014explaining} were based 
on perturbing all pixels by a small amount. We discuss in Section~\ref{sec:discussion} the
impact of our defense with other crafting algorithms, but use the above algorithm to confirm the
analytical results presented in the preceding sections. These two
steps are repeated several times until the resulting sample $X^*$ is classified
in the target class $F(X^*)$. 

We stop the perturbation selection if the number of features
perturbed is larger than $112$. This is justified because larger perturbations
would be detectable by humans~\cite{NAS-186} or potential anomaly detection systems. 
This method was previously reported to achieve a
$97\%$ success rate when used to craft $90,000$ adversarial samples by
altering samples from the MNIST test set with an average distortion of $4.02\%$
of the input features~\cite{NAS-186}. We find that altering a maximum of $112$ 
features also yields a high 
adversarial success rate of $92.78\%$ on the CIFAR10 test set. Note that throughout this evaluation, we
use the number of features altered while producing adversarial samples to
compare them with original samples.


\begin{figure*}
\centering
\includegraphics[width=\textwidth]{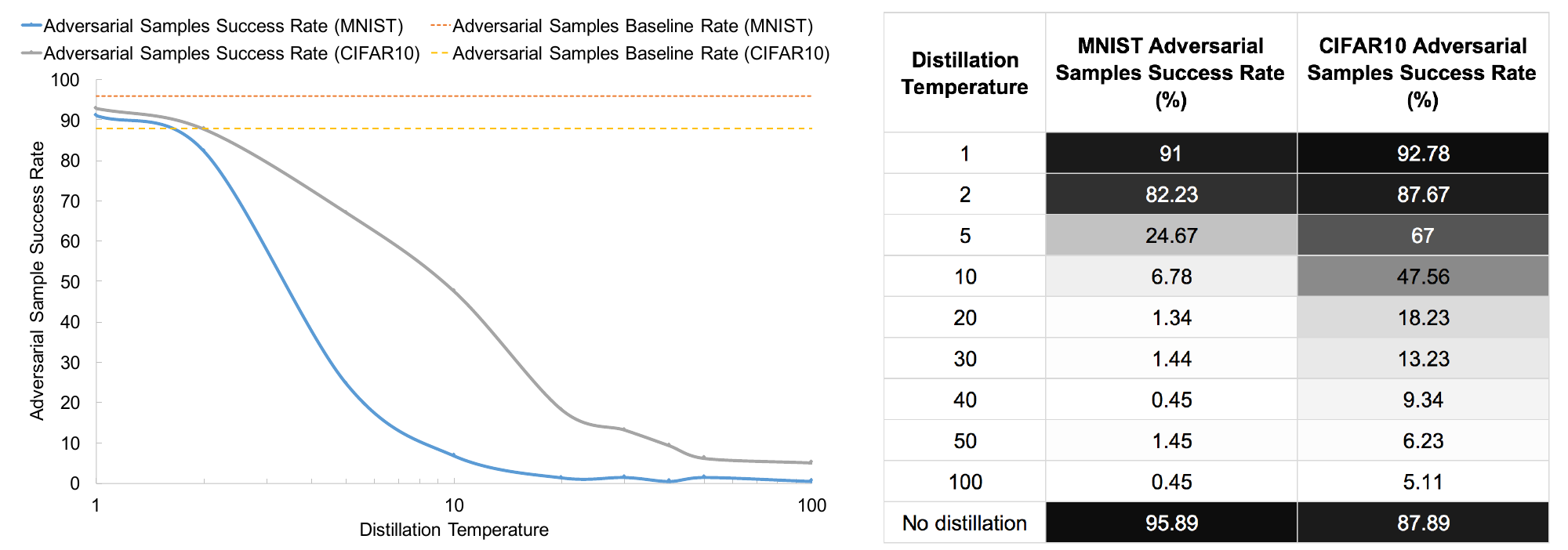}
\caption{\textbf{An exploration of the temperature parameter space:} for $900$
targets against the MNIST and CIFAR10 based models and several distillation
temperatures, we plot the percentage of targets achieved by crafting an
adversarial sample while altering at most $112$ features. Baselines for models
trained without distillation are in dashes. Note the horizontal logarithmic
scale.}
\label{fig:temperature-parameter-space}
\end{figure*}

\subsection{Defensive Distillation and Adversarial Samples}
\label{sec:defdistill}

\textbf{Impact on Adversarial Crafting} - For each of our two DNN architectures
corresponding to the MNIST and CIFAR10 datasets, we consider the original
trained model $F_{MNIST}$ or $F_{CIFAR10}$, as well as the distilled model
$F^d_{MNIST}$ or $F^d_{CIFAR10}$. We obtain the two distilled models by
training them with defensive distillation at a class knowledge transfer
temperature of $T=20$ (the choice of this parameter is investigated below). The
resulting classification accuracy for the MNIST model $F^d_{MNIST}$ is
$99.05\%$ and the classification accuracy for the CIFAR10 model $F^d_{CIFAR10}$
is $81.39\%$, which are comparable to the non-distilled models. 

In a second set of experiments, we measured success rate of adversarial sample
crafting on $100$ samples randomly selected from each dataset\footnote{Note
that we extract samples from the test set for convenience, but any sample
accepted as a network input could be used as the original sample.}. That is,
for each considered sample, we use the crafting algorithm to craft $9$
adversarial samples corresponding to the $9$ classes distinct from the sample'
source class. We thus craft a total of $900$ samples for each model. For the
architectures trained on MNIST data, we find that using defensive distillation
reduces the success rate of adversarial sample crafting from $95.89\%$ for the
original model to $1.34\%$ for the distilled model, thus resulting in a
$98.6\%$ decrease. Similarly, for the models trained on CIFAR10 data, we find
that using distillation reduces the success rate of adversarial sample crafting
from $89.9\%$ for the original model to $16.76\%$ for the distilled model,
which represents a $81.36\%$ decrease.

\textbf{Distillation Temperature} - The next experiments measure how
temperature impacts adversarial sample generation. Note the softmax layer's
temperature is set to $1$ at test time i.e., temperature only matters during
training.  The objective here is to identify the ``optimal'' training
temperature resulting in resilience to adversarial samples for a DNN and
dataset. 

We repeat the adversarial sample crafting experiment on both architectures and
vary the distillation temperature each time.  The number of adversarial targets
successfully reached for the following distillation temperatures $T$:
$\{1,2,5,10,20,30,50,100\}$ is measured.
Figure~\ref{fig:temperature-parameter-space} plots the success rate of
adversarial samples with respect to temperature for both architectures and
provides exact figures. In other words, the rate plotted is the number of
adversarial sample targets that were reached. Two interesting observations can
be made: (1) increasing the temperature will generally speaking make
adversarial sample crafting harder, and (2) there is an elbow point after which
the rate largely remains constant ($\approx 0\%$ for MNIST and $\approx 5\%$
for CIFAR10). 

Observation (1) validates analytical results from Section~\ref{sec:defense} showing distilled network
resilience to adversarial samples:
the success rate of adversarial crafting is reduced from $95.89\%$ without distillation to $0.45\%$
with distillation ($T=100$) on the MNIST based DNN, and from $87.89\%$ without distillation to $5.11\%$ with distillation ($T=100$) on the CIFAR10 DNN. 

The temperature corresponding to the curve elbow is linked to the role
temperature plays within the softmax layer. Indeed, temperature is used to
divide logits given as inputs to the softmax layer, in order to provide more
discreet or smoother distributions of probabilities for classes. Thus, one can
make the hypothesis that the curve's elbow is reached when the temperature is
such that increasing it further would not make the distribution smoother
because probabilities are already close to $1/N$ where $N$ is the number of
classes.  We confirm this hypothesis by computing the average maximum
probability output by the CIFAR10 DNN: it is equal to $0.72$ for $T=1$, to
$0.14$ for $T=20$, and to $0.11$ for $T=40$. Thus, the elbow point at $T=40$
correspond to probabilities near $1/N=0.1$. 

\begin{figure}
\centering
\includegraphics[width=\columnwidth]{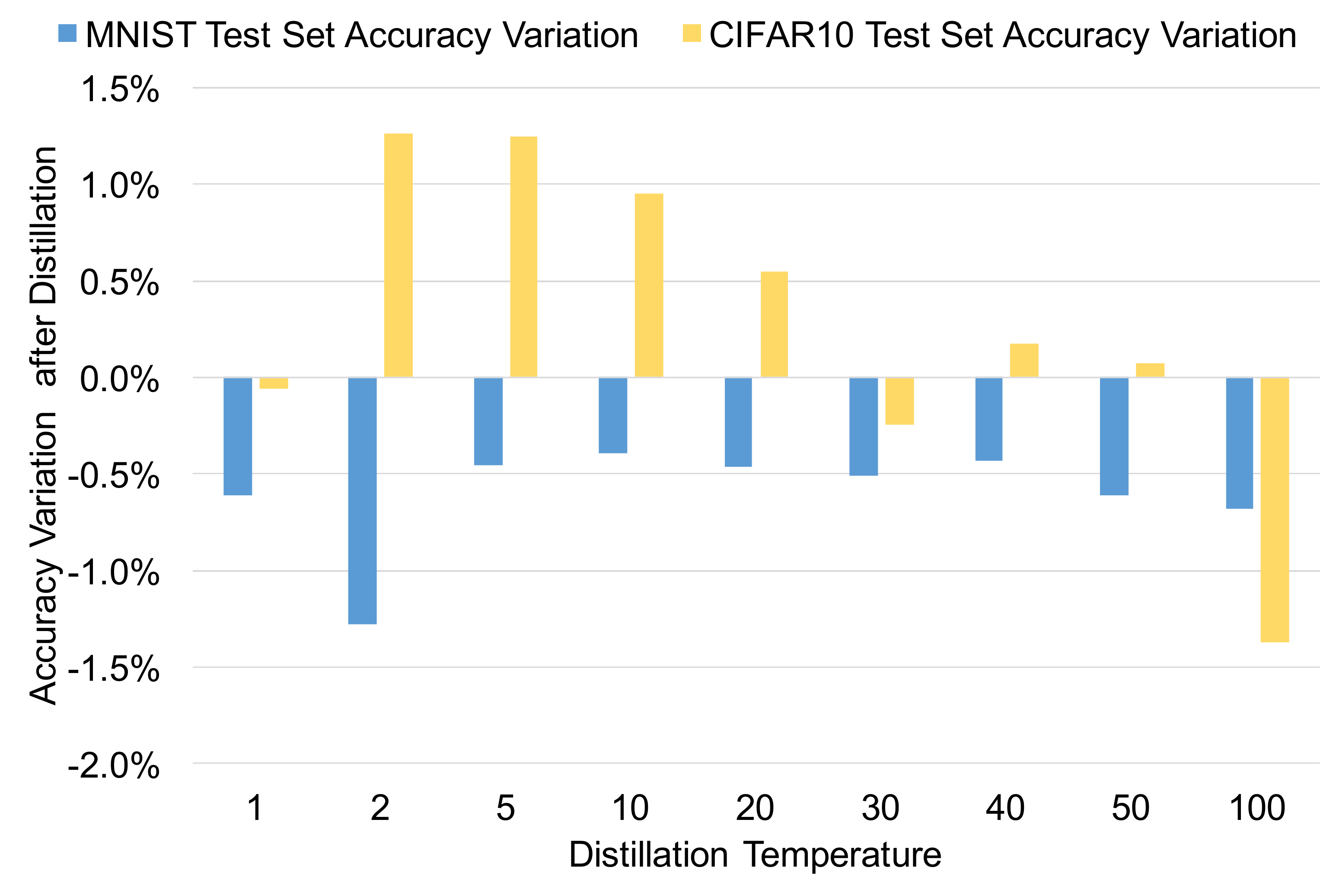}
\caption{\textbf{Influence of distillation on accuracy:} we plot the accuracy variations of our two architectures for a training with and without defensive distillation. These rates were evaluated on the corresponding test set for various temperature values.}
\label{fig:net-accuracies}
\end{figure}

\textbf{Classification Accuracy} - The next set of experiments sought to measure the impact of the approach on accuracy.
For each knowledge transfer temperature $T$ used in the previous set of
experiments, we compute the variation of classification accuracy between the
models $F_{MNIST}, F_{CIFAR10}$ and  $F^d_{MNIST}, F^d_{CIFAR10}$, respectively
trained without distillation and with distillation at temperature $T$. For each
model, the accuracy is computed using all $10,000$ samples from the
corresponding test set (from MNIST for the first and from CIFAR10 for the
second model). Recall that the baseline rate, meaning the accuracy rate
corresponding to training performed without distillation, which we computed
previously was $99.51\%$ for model $F_{MNIST}$ and $80.95\%$ for model
$F_{CIFAR10}$. The variation rates for the set of distillation temperatures are
shown in Figure~\ref{fig:net-accuracies}.  

\begin{figure*}[t]
\centering
\includegraphics[width=\textwidth]{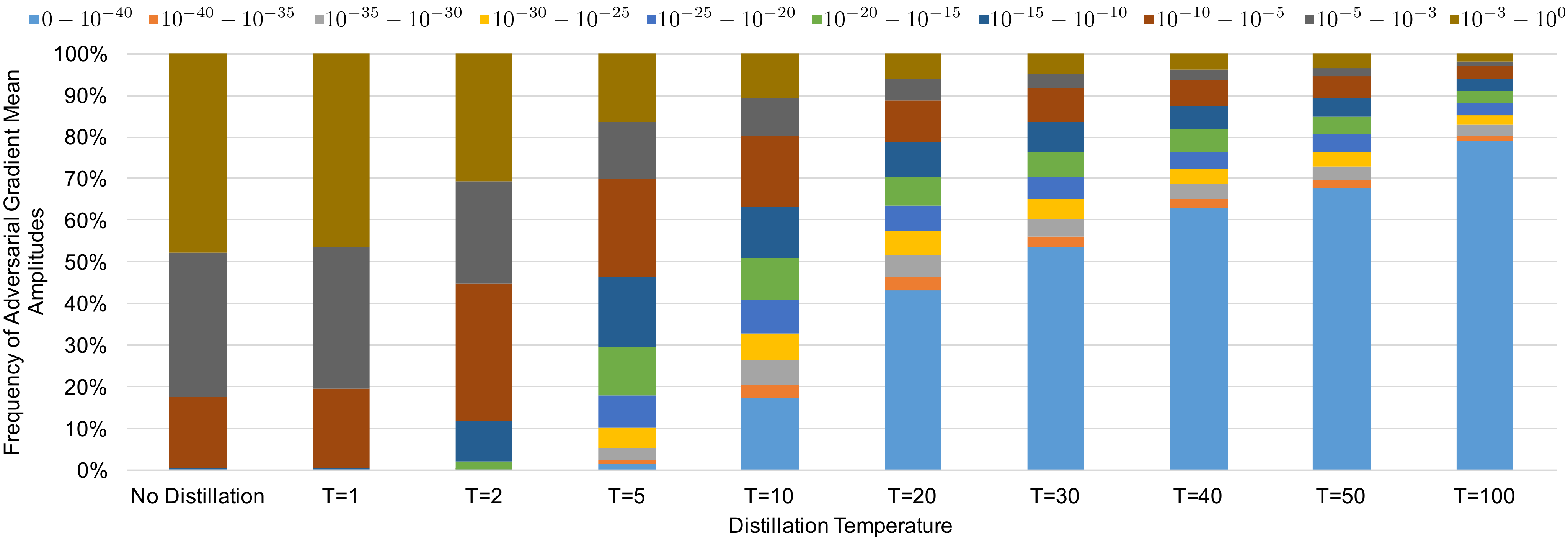}
\caption{\textbf{An exploration of the impact of temperature on the amplitude
of adversarial gradients:} We illustrate how adversarial gradients vanish as
distillation is performed at higher temperatures. Indeed, for each temperature
considered, we draw the repartition of samples in each of the 10 ranges of mean
adversarial gradient amplitudes associated with a distinct color. This data was
collected using all $10,000$ samples from the CIFAR10 test set on the
corresponding DNN model.}
\label{fig:temperature-impact-on-fwder}
\end{figure*}

One can observe that variations in accuracy introduced by distillation are moderate. 
 For instance, the accuracy of the MNIST based model is
degraded by less than $1.28\%$ for all temperatures, with for instance 
an accuracy of $99.05\%$ for $T=20$, which would have been state of the art
until very recently.  Similarly, the accuracy of the
CIFAR10 based model is degraded by at most $1.37\%$. It also potentially
improves it, as some variations are positive, notably for the CIFAR10 model
(the MNIST model is hard to improve because its accuracy is already close to a
$100\%$). Although providing a quantitative understanding of this potential
for accuracy improvement is outside the scope of this paper, we believe that it
stems from the generalization capabilities favored by distillation, as
investigated in the analytical study of defensive distillation conducted
previously in Section~\ref{sec:defense}. 
 
To summarize, not only distillation improves resilience of DNNs to adversarial
perturbations (from $95.89\%$ to $0.45\%$ on a first DNN, and from $87.89\%$ to
$5.11\%$ on a second DNN), it also does so without severely impacting
classification correctness (the accuracy variability between models trained
without distillation and with distillation is smaller than $1.37\%$ for both
DNNs). Thus, defensive distillation matches the second
defense requirement from Section~\ref{sec:adversarial-deep-learning}.
When deploying defensive distillation, defenders will have to empirically 
find a temperature value $T$ offering a good balance between robustness
to adversarial perturbations and classification accuracy. In our case, for 
the MNIST model for instance, such a temperature would be $T=20$ according
 to Figure~\ref{fig:temperature-parameter-space} and~\ref{fig:net-accuracies}.


\vspace*{-0.2in}

\subsection{Distillation and Sensitivity}
\label{sec:sensitivity}

The second battery of experiments sought to demonstrate the impact of
distillation on a DNN's sensitivity to inputs. Our hypothesis is that our
defense mechanism reduces gradients exploited by adversaries to craft
perturbations. To confirm this hypothesis, we evaluate the mean amplitude of
these gradients on models trained without and with defensive distillation.  In
this experiment, we split the $10,000$ samples from the CIFAR10 test set into
bins according to the mean value of their adversarial gradient amplitude.  We
train these at varying temperatures and plot the resulting bin frequencies in
Figure~\ref{fig:temperature-impact-on-fwder}.

Note that distillation reduces the average absolute value of adversarial
gradients: for instance the mean adversarial gradient amplitude without
distillation is larger than $0.001$ for 4763 samples among the 10,000 samples
considered, whereas it is the case only for 172 samples when distillation is
performed at a temperature of $T=100$. Similarly, 8 samples are in the bin
corresponding to a mean adversarial gradient amplitude smaller than $10^{-40}$
for the model trained without distillation, whereas there is a vast majority of
samples, namely 7908 samples, with a mean adversarial gradient amplitude
smaller than $10^{-40}$ for the model trained with defensive distillation at a
temperature of $T=100$. Generally speaking one can observe that the largest
frequencies of samples shifts from higher mean amplitudes of adversarial
gradients to smaller ones.

When the amplitude of adversarial gradients is smaller, it means the DNN model
learned during training is smoother around points in the distribution
considered. This in turns means that evaluating the sensitivity of directions
will be more complex and crafting adversarial samples will require adversaries
to introduce more perturbation for the same original samples. Another
observation is that overtraining does not help because when there is
overfitting, the adversarial gradients progressively increase in amplitude so
early stopping and other similar techniques can help to prevent exploding. This
is further discussed in Section~\ref{sec:discussion}. In our case, training for
50 epochs was sufficient for distilled DNN models to achieve comparable
accuracies to original models, and ensured that adversarial gradients did not
explode.    These experiments show that  distillation can have a smoothing
impact on classification models learned during training. Indeed, gradients
characterizing model sensitivity to input variations are reduced by factors
larger than $10^{30}$ when defensive distillation is applied.

\vspace*{-0.1in}

\subsection{Distillation and Robustness}
\label{sec:robust}

Lastly, we explore the interplay between smoothness of classifiers and
robustness.  Intuitively, robustness is the average minimal perturbation
required to produce an adversarial sample from the distribution modeled by $F$. 

\begin{figure}[t]
\centering
\includegraphics[width=\columnwidth]{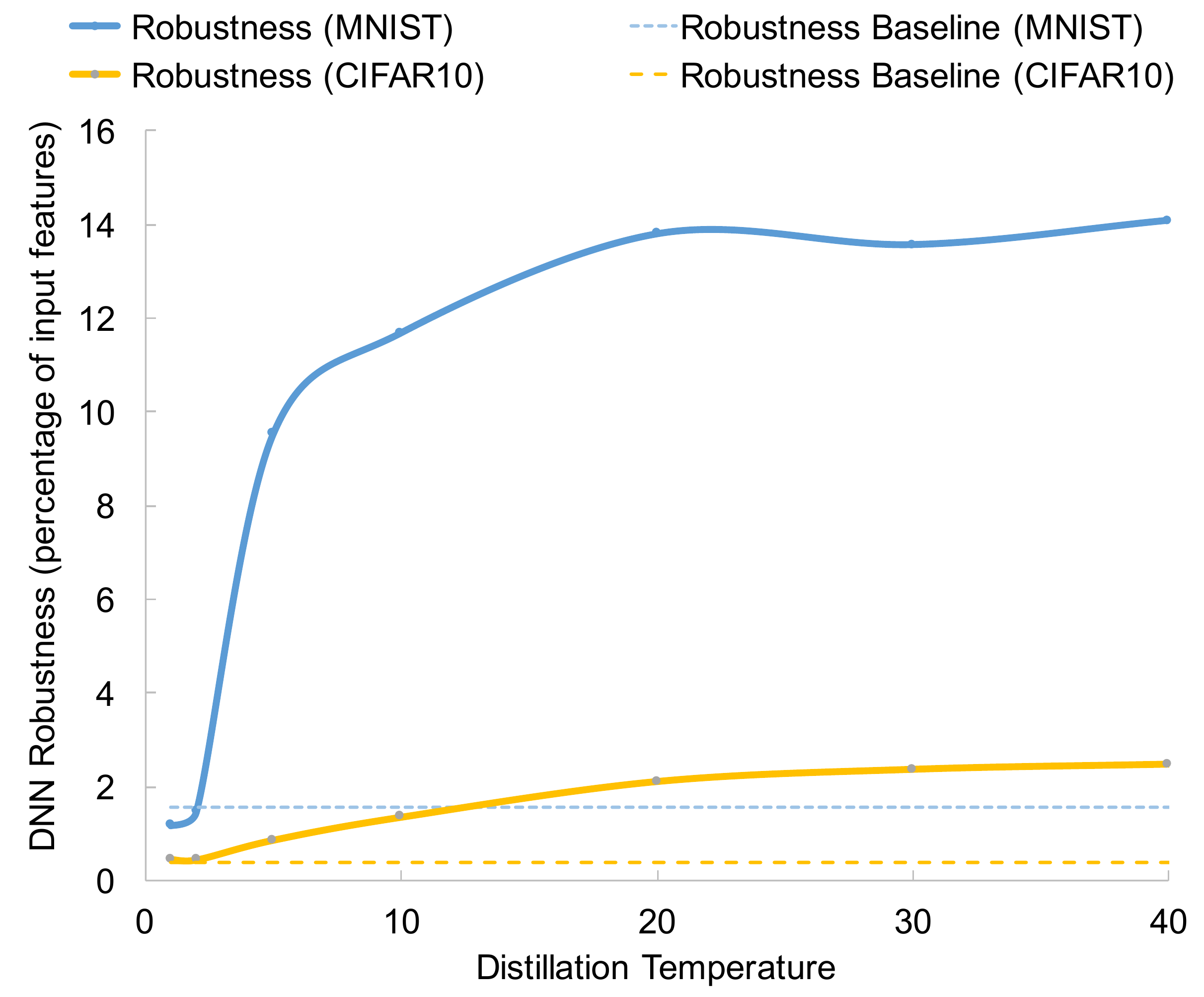}
\caption{\textbf{Quantifying the impact of distillation temperature on
robustness:} we plot the value of robustness described in
Equation~\ref{eq:exp-robustness} for several temperatures and compare it to a
baseline robustness value for models trained without distillation.}
\label{fig:temperature-robustness}
\end{figure}

\textbf{Robustness} -  Recall the definition of robustness: 
\begin{equation}
\rho_{adv}(F)=E_{\mu}[\Delta_{adv}(X,F)]
\end{equation} 
where inputs $X$ are drawn from distribution $\mu$ that DNN architecture $F$ is trying to model, and $\Delta_{adv}(X,F)$ is defined  in Equation~\ref{eq:min-misclassification-perturbation} to be the minimum perturbation required to misclassify sample $X$ in each of the other classes.
We now evaluate whether distillation effectively increases this robustness
metric for our evaluation architectures. To do this without exhaustively
searching all perturbations for each possible sample of the underlying
distribution modeled by the DNN, we approximate the metric: we compute the
metric over all 10,000 samples in the test set for each model.  This results in
the computation of the following quantity: 
\begin{equation}
\label{eq:exp-robustness}
\rho_{adv}(F)\simeq \frac{1}{|\cal{X}|}\sum_{X\in\cal{X}} \min_{\delta X} \|\delta X\|
\end{equation} 
where values of $\delta X$ are evaluated by considering each of the 9 possible
adversarial targets corresponding to sample $X\in \cal{X}$. and using the
number of features altered while creating the corresponding adversarial samples
as the distance measure to evaluate the minimum perturbation $\|\delta X\|$
required to create the mentionned adversarial sample. In
Figure~\ref{fig:temperature-robustness}, we plot the evolution of the
robustness metric with respect to an increase in distillation temperature for
both architectures. One can see that as the temperature increases, the
robustness of the network as defined here, increases. For the MNIST
architecture, the model trained without distillation displays a robustness of
$1.55\%$ whereas the model trained with distillation at $T=20$ displays a
robustness of $13.79\%$, an increase of $790\%$. Note that, perturbations of $13.79\%$
are large enough that they potentially change the true class or could
be detected by an anomaly detection process. In fact, it 
was empirically shown in previous work that humans 
begin to misclassify adversarial samples (or at least identify them as erroneous)
for perturbations larger than $13.79\%$: see Figure 16 in~\cite{NAS-186}. It is not desirable for adversary 
to produce adversarial samples identifiable by humans. Furthermore, changing additional features 
can be hard, depending on the input nature. In this evaluation, it is easy to change
 a feature in the images. However, if the input was spam email, it would become 
 challenging for the adversary to alter many input features. Thus, making DNNs robust
  to small perturbations is of paramount importance. 
 Similarly, for the CIFAR10
architecture, the model trained without distillations displays a robustness of
$0.39\%$ whereas the model trained with defensive distillation at $T=50$ has a
robustness of $2.56\%$, which represents an increase of $556\%$. This result
suggests that indeed distillation is able to provide sufficient additional
knowledge to improve the generalization capabilities of DNNs outside of their
training manifold, thus developing their robustness to  perturbations.

\textbf{Distillation and Confidence -} Next we investigate the impact of
distillation temperature on DNN classification confidence.  Our hypothesis is
that distillation also impacts the confidence of class predictions made by
distilled model. 

To test this hypothesis, we evaluate the confidence prediction for all $10,000$
samples in the CIFAR10 dataset. We average the following quantity over all
samples $X\in \cal{X}$: 
\begin{equation}
C(X) = \left\lbrace
\begin{array}{c}
0  \mbox{ if } \arg\max_{i} F_i(X) \neq t(X)\\
 \arg\max_{i} F_i(X) \mbox{ otherwise}
\end{array}\right.
\end{equation}
where $t(X)$ is the correct class of sample $X$. The resulting confidence
values are shown in Table~\ref{fig:t-confidence} where the lowest confidence is
$0\%$ and the highest $100\%$. The monotonically increasing trend suggests that
distillation does indeed increase predictive confidence.  Note that a similar
analysis of MNIST is inconclusive because all confidence values are already
near $99\%$, which leaves little opportunity for improvement.

\begin{table}[h]
\centering
\begin{tabular}{|c|c|c|c|c|c|}
\hline
 $T$ & 1 & 2 & 5 & 10 & 20  \\ \hline 
$C(X)$ & 71.85\% & 71.99\% & 78.05\% & 80.77\%  & 81.06\%   \\ \hline 
\end{tabular}
\caption{\textbf{CIFAR10 model prediction confidence:} $C(X)$ is evaluated on the test set at various temperatures $T$. }
\label{fig:t-confidence}
\end{table}


\section{Discussion}
\label{sec:discussion}

The preceding analysis of distillation shows that it can increase the
resilience of DNNs to adversarial samples. Training extracts knowledge learned
about classes from probability vectors produced by the DNN. Resulting models
have stronger generalizations capabilities outside of their training set.

A limitation of defensive distillation is that it is only applicable to DNN
models that produce an energy-based probability distribution, for which 
a temperature can be defined. Indeed, this paper's implementation of distillation is
dependent on an engergy-based probability distribution for two reasons: the softmax produces the
probability vectors and introduces the temperature parameter.
Thus, using defensive distillation in machine learning models different from DNNs would require
additional research efforts. However note that many machine learning
 models, unlike DNNs, don't have the model capacity to be able to resist adversarial examples. 
 For instance, Goodfellow et al.~\cite{goodfellow2014explaining} showed that
 shallow models like linear models are also vulnerable to adversarial examples
 and are unlikely to be hardened against them.
A defense specialized to DNNs, guaranteed by the universal approximation property 
to at least be able to represent a function that correctly processes adversarial examples, is thus a significant
step towards building machine learning models robust to adversarial samples. 

In our evaluation setup, we defined the distance measure between original
samples and adversarial samples as the number of modified features. There are
other metrics suitable to compare samples, like $L1, L2$ norms. Using different
metrics will produce different distortions and can be pertinent in application
domains different from computer vision. For instance, crafting adversarial
samples from real malware to evade existing detection methods will require
different metrics and perturbations~\cite{fogla2006evading,
biggio2014poisoning}. Future work should investigate the use of various
distance measures. 

One question is also whether the probabilities, used to transfer knowledge in this paper, could be replaced by soft class labels. For a $N$-class
classification problem, soft labels are obtained by replacing the target value of $1$ for the correct class with a
 target value of $0.9$, and for the incorrect classes replacing the target of 0 with $\frac{1}{10\cdot N}$.
We empirically observed that the improvements to the neural network's robustness are not as significant with soft labels. 
Specifically, we trained the MNIST DNN used
in Section~\ref{sec:evaluation} using soft labels. The misclassification rate of
adversarial samples, crafted using MNIST test data and the same attack 
parameters than in Section~\ref{sec:evaluation}, was of $86.00\%$, whereas the 
distilled model studied in Section~\ref{sec:evaluation} had a misclassification rate smaller than $1\%$ We believe this is due to the relative information between classes encoded in probability vectors and not in soft class labels.
Inspired by an early public preprint of this paper, Warde-Farley and 
 Goodfellow~\cite{WardeFarley16}  independently tested label smoothing, and found that it 
 partially resists adversarial examples crafted using the fast gradient sign method~\cite{goodfellow2014explaining}.
 One possible interpretation of these conflicting results is that label smoothing 
 without distillation is smart enough to defend against simple, inexpensive 
 methods~\cite{goodfellow2014explaining} of adversarial example crafting but not more powerful iterative methods used in this paper~\cite{NAS-186}. 
  
Future work should also evaluate the performance of defensive distillation in the face of
different perturbation types. For instance, while defensive distillation is a good 
defense against the attack studied here~\cite{NAS-186}, it could still be vulnerable to other 
attacks based on L-BFGS~\cite{szegedy2013intriguing}, the fast gradient sign method~\cite{goodfellow2014explaining}, or genetic algorithms~\cite{nguyen2014deep}. 
However, against such techniques, the preliminary results from~\cite{WardeFarley16}
 are promising and worthy of exploration; it seems likely that
distillation will also have a beneficial defensive impact with such techniques. 
  
In this paper, we did not compare our defense technique to traditional regularization
techniques because adversarial examples are not a traditional overfitting 
problem~\cite{goodfellow2014explaining}. In fact, previous work showed that
a wide variety of traditional regularization methods including dropout and weight decay
either fail to defend against adversarial examples or only do so by
 seriously harming accuracy on the original task~\cite{szegedy2013intriguing,goodfellow2014explaining}.

Finally, we would like to point out that defensive distillation does 
not create additional attack vectors, in other words does not start an
 arms race between defenders and attackers. Indeed, the attacks~\cite{szegedy2013intriguing,goodfellow2014explaining,NAS-186} are designed 
 to be approximately optimal regardless of the targeted model. Even if an attacker 
knows that defensive distillation is being used, it is not clear how he could 
exploit this to adapt its attack. By increasing  confidence estimates across a lot of the model's
 input space, defensive distillation should lead to strictly better models.


\section{Related Work}
\label{sec:related-work}

Machine learning security~\cite{barreno2010security} is an active
research area in the security community~\cite{xuautomatically}. Attacks have been
organized in taxonomies according to adversarial capabilities 
in~\cite{huang2011adversarial, barreno2006can}. Biggio et al. studied binary classifiers deployed in adversarial
settings and proposed a framework to secure them~\cite{biggio2014security}.
Their work does not consider deep learning models but rather binary classifiers
 like Support Vector Machines or logistic regression.
More generally, attacks against machine learning models can be 
partitioned by execution time: during training~\cite{biggio2012poisoning,biggio2011support} or 
at test time~\cite{biggio2013evasion} when the model is used to make predictions. 

Previous work studying DNNs in adversarial settings focused on presenting novel
attacks against DNNs at test time, mainly exploiting vulnerabilities to
adversarial samples~\cite{NAS-186, goodfellow2014explaining,
szegedy2013intriguing}. These attacks were discussed in depth in
section~\ref{sec:adversarial-deep-learning}. These papers offered suggestions
for defenses but their investigation was left to future work by all authors,
whereas we proposed and evaluated a full defense mechanism to improve the
resilience of DNNs to adversarial perturbations. 

Nevertheless some attempts were made at making DNN resilient to adversarial
perturbations. Goodfellow et al. showed that radial basis activation functions
are more resistant to perturbations, but deploying them requires important
modifications to the existing architecture~\cite{goodfellow2014explaining}. Gu
et al. explored the use of denoising auto-encoders, a DNN type of architecture
intended to capture main factors of variation in the data, and showed that they
can remove substantial amounts of adversarial noise~\cite{gu2014towards}.
However the resulting stacked architecture can again be evaded using
adversarial samples. The authors therefore proposed a new architecture, Deep
Contractive Networks, based on imposing layer-wise penalty defined using the
network's Jacobian. This penalty however limits the capacity of Deep
Contractive Networks compared to traditional DNNs. 


\section{Conclusions}
\label{sec:conclusion}

In this work we have investigated the use of distillation, a technique
previously used to reduce DNN dimensionality, as a defense against adversarial
perturbations.  We formally defined \emph{defensive distillation} and evaluated
it on standard DNN architectures. Using elements of learning theory, we
analytically showed how distillation impacts models learned by deep neural
network architectures during training. Our empirical findings show that
\emph{defensive distillation} can significantly reduce the successfulness of
attacks against DNNs. It reduces the success of adversarial sample crafting to
rates smaller than $0.5\%$ on the MNIST dataset and smaller than $5\%$ on the
CIFAR10 dataset while maintaining the accuracy rates of the original DNNs.
Surprisingly, distillation is simple to implement and introduces very little
overhead during training. Hence, this work lays out a new foundation for
securing systems based on deep learning. 

Future work should investigate the impact of distillation on other DNN
models and adversarial sample crafting algorithms. One notable endeavor is to extend this
approach outside of the scope of classification to other DL tasks. This is not
trivial as it requires finding a substitute for probability vectors used in
defensive distillation with similar properties. Lastly, we will explore
different definitions of robustness that measure other aspects of DNN
resilience to adversarial perturbations.


\section*{Acknowledgment}
The authors would like to thank Damien Octeau, Ian Goodfellow, and Ulfar Erlingsson for their insightful comments. Research was sponsored by the Army Research Laboratory and was accomplished
under Cooperative Agreement Number W911NF-13-2-0045 (ARL Cyber Security
CRA). The views and conclusions contained in this document are those of the
authors and should not be interpreted as representing the official policies,
either expressed or implied, of the Army Research Laboratory or the U.S.
Government. The U.S. Government is authorized to reproduce and distribute
reprints for Government purposes notwithstanding any copyright notation here
on.




\begin{thebibliography}{10}
\providecommand{\url}[1]{#1}
\csname url@samestyle\endcsname
\providecommand{\newblock}{\relax}
\providecommand{\bibinfo}[2]{#2}
\providecommand{\BIBentrySTDinterwordspacing}{\spaceskip=0pt\relax}
\providecommand{\BIBentryALTinterwordstretchfactor}{4}
\providecommand{\BIBentryALTinterwordspacing}{\spaceskip=\fontdimen2\font plus
\BIBentryALTinterwordstretchfactor\fontdimen3\font minus
  \fontdimen4\font\relax}
\providecommand{\BIBforeignlanguage}[2]{{%
\expandafter\ifx\csname l@#1\endcsname\relax
\typeout{** WARNING: IEEEtran.bst: No hyphenation pattern has been}%
\typeout{** loaded for the language `#1'. Using the pattern for}%
\typeout{** the default language instead.}%
\else
\language=\csname l@#1\endcsname
\fi
#2}}
\providecommand{\BIBdecl}{\relax}
\BIBdecl

\bibitem{krizhevsky2012imagenet}
A.~Krizhevsky, I.~Sutskever, and G.~E. Hinton, ``Imagenet classification with
  deep convolutional neural networks,'' in \emph{Advances in neural information
  processing systems}, 2012, pp. 1097--1105.

\bibitem{sainath2013deep}
T.~N. Sainath, A.-r. Mohamed, B.~Kingsbury, and B.~Ramabhadran, ``Deep
  convolutional neural networks for lvcsr,'' in \emph{Acoustics, Speech and
  Signal Processing (ICASSP), 2013 IEEE International Conference on}.\hskip 1em
  plus 0.5em minus 0.4em\relax IEEE, 2013, pp. 8614--8618.

\bibitem{sermanet2014overfeat}
P.~Sermanet, D.~Eigen, X.~Zhang, M.~Mathieu, R.~Fergus, and Y.~LeCun,
  ``Overfeat: Integrated recognition, localization and detection using
  convolutional networks,'' in \emph{International Conference on Learning
  Representations (ICLR 2014)}.\hskip 1em plus 0.5em minus 0.4em\relax arXiv
  preprint arXiv:1312.6229, 2014.

\bibitem{dahl2013large}
G.~E. Dahl, J.~W. Stokes, L.~Deng, and D.~Yu, ``Large-scale malware
  classification using random projections and neural networks,'' in
  \emph{Acoustics, Speech and Signal Processing (ICASSP), 2013 IEEE
  International Conference on}.\hskip 1em plus 0.5em minus 0.4em\relax IEEE,
  2013, pp. 3422--3426.

\bibitem{yuan2014droid}
Z.~Yuan, Y.~Lu, Z.~Wang, and Y.~Xue, ``Droid-sec: deep learning in android
  malware detection,'' in \emph{Proceedings of the 2014 ACM conference on
  SIGCOMM}.\hskip 1em plus 0.5em minus 0.4em\relax ACM, 2014, pp. 371--372.

\bibitem{PayPal}
\BIBentryALTinterwordspacing
E.~Knorr, ``How paypal beats the bad guys with machine learning,'' 2015.
  [Online]. Available:
  \url{http://www.infoworld.com/article/2907877/machine-learning/how-paypal-reduces-fraud-with-machine-learning.html}
\BIBentrySTDinterwordspacing

\bibitem{NAS-186}
N.~Papernot, P.~McDaniel, S.~Jha, M.~Fredrikson, Z.~B. Celik, and A.~Swami,
  ``The limitations of deep learning in adversarial settings,'' in
  \emph{Proceedings of the 1st IEEE European Symposium on Security and
  Privacy}.\hskip 1em plus 0.5em minus 0.4em\relax IEEE, 2016.

\bibitem{szegedy2013intriguing}
C.~Szegedy, W.~Zaremba, I.~Sutskever, J.~Bruna, D.~Erhan, I.~Goodfellow, and
  R.~Fergus, ``Intriguing properties of neural networks,'' in \emph{Proceedings
  of the 2014 International Conference on Learning Representations}.\hskip 1em
  plus 0.5em minus 0.4em\relax Computational and Biological Learning Society,
  2014.

\bibitem{goodfellow2014explaining}
I.~J. Goodfellow, J.~Shlens, and C.~Szegedy, ``Explaining and harnessing
  adversarial examples,'' in \emph{Proceedings of the 2015 International
  Conference on Learning Representations}.\hskip 1em plus 0.5em minus
  0.4em\relax Computational and Biological Learning Society, 2015.

\bibitem{NVIDIATegra}
\BIBentryALTinterwordspacing
NVIDIA, ``Nvidia tegra drive px: Self-driving car computer,'' 2015. [Online].
  Available: \url{http://www.nvidia.com/object/drive-px.html}
\BIBentrySTDinterwordspacing

\bibitem{cirecsan2012multi}
D.~Cire{\c{s}}an, U.~Meier, J.~Masci \emph{et~al.}, ``Multi-column deep neural
  network for traffic sign classification.''

\bibitem{huang2011adversarial}
L.~Huang, A.~D. Joseph, B.~Nelson, B.~I. Rubinstein, and J.~Tygar,
  ``Adversarial machine learning,'' in \emph{Proceedings of the 4th ACM
  workshop on Security and artificial intelligence}.\hskip 1em plus 0.5em minus
  0.4em\relax ACM, 2011, pp. 43--58.

\bibitem{biggio2014pattern}
B.~Biggio, G.~Fumera \emph{et~al.}, ``Pattern recognition systems under attack:
  Design issues and research challenges,'' \emph{International Journal of
  Pattern Recognition and Artificial Intelligence}, vol.~28, no.~07, p.
  1460002, 2014.

\bibitem{biggio2013evasion}
B.~Biggio, I.~Corona, D.~Maiorca, B.~Nelson \emph{et~al.}, ``Evasion attacks
  against machine learning at test time,'' in \emph{Machine Learning and
  Knowledge Discovery in Databases}.\hskip 1em plus 0.5em minus 0.4em\relax
  Springer, 2013, pp. 387--402.

\bibitem{anjos2011counter}
A.~Anjos and S.~Marcel, ``Counter-measures to photo attacks in face
  recognition: a public database and a baseline,'' in \emph{Proceedings of the
  2011 International Joint Conference on Biometrics}.\hskip 1em plus 0.5em
  minus 0.4em\relax IEEE, 2011.

\bibitem{fogla2006evading}
P.~Fogla and W.~Lee, ``Evading network anomaly detection systems: formal
  reasoning and practical techniques,'' in \emph{Proceedings of the 13th ACM
  conference on Computer and communications security}.\hskip 1em plus 0.5em
  minus 0.4em\relax ACM, 2006, pp. 59--68.

\bibitem{gu2014towards}
S.~Gu and L.~Rigazio, ``Towards deep neural network architectures robust to
  adversarial examples,'' in \emph{Proceedings of the 2015 International
  Conference on Learning Representations}.\hskip 1em plus 0.5em minus
  0.4em\relax Computational and Biological Learning Society, 2015.

\bibitem{ba2014deep}
J.~Ba and R.~Caruana, ``Do deep nets really need to be deep?'' in
  \emph{Advances in Neural Information Processing Systems}, 2014, pp.
  2654--2662.

\bibitem{hinton2015distilling}
G.~Hinton, O.~Vinyals, and J.~Dean, ``Distilling the knowledge in a neural
  network,'' in \emph{Deep Learning and Representation Learning Workshop at
  NIPS 2014}.\hskip 1em plus 0.5em minus 0.4em\relax arXiv preprint
  arXiv:1503.02531, 2014.

\bibitem{lecun1998mnist}
Y.~LeCun and C.~Cortes, ``The mnist database of handwritten digits,'' 1998.

\bibitem{krizhevsky2009learning}
A.~Krizhevsky and G.~Hinton, ``Learning multiple layers of features from tiny
  images,'' 2009.

\bibitem{Bengio-et-al-2015-Book}
\BIBentryALTinterwordspacing
Y.~Bengio, I.~J. Goodfellow, and A.~Courville, ``Deep learning,'' 2015, book in
  preparation for MIT Press. [Online]. Available:
  \url{http://www.iro.umontreal.ca/\textasciitilde bengioy/dlbook}
\BIBentrySTDinterwordspacing

\bibitem{hinton2007learning}
G.~E. Hinton, ``Learning multiple layers of representation,'' \emph{Trends in
  cognitive sciences}, vol.~11, no.~10, pp. 428--434, 2007.

\bibitem{rumelhart1988learning}
D.~E. Rumelhart, G.~E. Hinton, and R.~J. Williams, ``Learning representations
  by back-propagating errors,'' \emph{Cognitive modeling}, vol.~5, 1988.

\bibitem{bergstra2012random}
J.~Bergstra and Y.~Bengio, ``Random search for hyper-parameter optimization,''
  \emph{The Journal of Machine Learning Research}, vol.~13, no.~1, pp.
  281--305, 2012.

\bibitem{glorot2011domain}
X.~Glorot, A.~Bordes, and Y.~Bengio, ``Domain adaptation for large-scale
  sentiment classification: A deep learning approach,'' in \emph{Proceedings of
  the 28th International Conference on Machine Learning (ICML-11)}, 2011, pp.
  513--520.

\bibitem{masci2011stacked}
J.~Masci, U.~Meier, D.~Cire{\c{s}}an \emph{et~al.}, ``Stacked convolutional
  auto-encoders for hierarchical feature extraction,'' in \emph{Artificial
  Neural Networks and Machine Learning--ICANN 2011}.\hskip 1em plus 0.5em minus
  0.4em\relax Springer, 2011, pp. 52--59.

\bibitem{erhan2010does}
D.~Erhan, Y.~Bengio, A.~Courville, P.-A. Manzagol, P.~Vincent, and S.~Bengio,
  ``Why does unsupervised pre-training help deep learning?'' \emph{The Journal
  of Machine Learning Research}, vol.~11, pp. 625--660, 2010.

\bibitem{miyato2015distributional}
T.~Miyato, S.~Maeda, M.~Koyama \emph{et~al.}, ``Distributional smoothing by
  virtual adversarial examples,'' \emph{CoRR}, vol. abs/1507.00677, 2015.

\bibitem{fawzi2015analysis}
A.~Fawzi, O.~Fawzi, and P.~Frossard, ``Analysis of classifiers' robustness to
  adversarial perturbations,'' in \emph{Deep Learning Workshop at ICML
  2015}.\hskip 1em plus 0.5em minus 0.4em\relax arXiv preprint
  arXiv:1502.02590, 2015.

\bibitem{drucker1992improving}
H.~Drucker and Y.~Le~Cun, ``Improving generalization performance using double
  backpropagation,'' \emph{Neural Networks, IEEE Transactions on}, vol.~3,
  no.~6, pp. 991--997, 1992.

\bibitem{nguyen2014deep}
A.~Nguyen, J.~Yosinski, and J.~Clune, ``Deep neural networks are easily fooled:
  High confidence predictions for unrecognizable images,'' in \emph{In Computer
  Vision and Pattern Recognition (CVPR 2015)}.\hskip 1em plus 0.5em minus
  0.4em\relax IEEE, 2015.

\bibitem{Cybenko92}
G.~Cybenko, ``Approximation by superpositions of a sigmoidal function,''
  \emph{{Mathematics of Control, Signals, and Systems }}, vol.~5, no.~4, p.
  455, 1992.

\bibitem{SSSSS10}
S.~Shalev-Shwartz, O.~Shamir, N.~Srebro, and K.~Sridharan, ``Learnability,
  stability and uniform convergence,'' \emph{The Journal of Machine Learning
  Research}, vol.~11, pp. 2635--2670, 2010.

\bibitem{bergstra2010theano}
J.~Bergstra, O.~Breuleux, F.~Bastien, P.~Lamblin, R.~Pascanu, G.~Desjardins,
  J.~Turian, D.~Warde-Farley, and Y.~Bengio, ``Theano: a cpu and gpu math
  expression compiler,'' in \emph{Proceedings of the Python for scientific
  computing conference (SciPy)}, vol.~4.\hskip 1em plus 0.5em minus 0.4em\relax
  Austin, TX, 2010, p.~3.

\bibitem{lasagne}
\BIBentryALTinterwordspacing
E.~Battenberg, S.~Dieleman, D.~Nouri, E.~Olson, A.~van~den Oord, C.~Raffel,
  J.~Schlüter, and S.~Kaae~Sønderby, ``Lasagne: Lightweight library to build
  and train neural networks in theano,'' 2015. [Online]. Available:
  \url{https://github.com/Lasagne/Lasagne}
\BIBentrySTDinterwordspacing

\bibitem{biggio2014poisoning}
B.~Biggio, K.~Rieck, D.~Ariu, C.~Wressnegger \emph{et~al.}, ``Poisoning
  behavioral malware clustering,'' in \emph{Proceedings of the 2014 Workshop on
  Artificial Intelligent and Security Workshop}.\hskip 1em plus 0.5em minus
  0.4em\relax ACM, 2014, pp. 27--36.

\bibitem{WardeFarley16}
D.~Warde-Farley and I.~Goodfellow, ``Adversarial perturbations of deep neural
  networks,'' in \emph{Advanced Structured Prediction}, T.~Hazan,
  G.~Papandreou, and D.~Tarlow, Eds., 2016.

\bibitem{barreno2010security}
M.~Barreno, B.~Nelson, A.~D. Joseph, and J.~Tygar, ``The security of machine
  learning,'' \emph{Machine Learning}, vol.~81, no.~2, pp. 121--148, 2010.

\bibitem{xuautomatically}
W.~Xu, Y.~Qi \emph{et~al.}, ``Automatically evading classifiers,'' in
  \emph{Proceedings of the 2016 Network and Distributed Systems Symposium},
  2016.

\bibitem{barreno2006can}
M.~Barreno, B.~Nelson, R.~Sears, A.~D. Joseph, and J.~D. Tygar, ``Can machine
  learning be secure?'' in \emph{Proceedings of the 2006 ACM Symposium on
  Information, computer and communications security}.\hskip 1em plus 0.5em
  minus 0.4em\relax ACM, 2006, pp. 16--25.

\bibitem{biggio2014security}
B.~Biggio, G.~Fumera, and F.~Roli, ``Security evaluation of pattern classifiers
  under attack,'' \emph{Knowledge and Data Engineering, IEEE Transactions on},
  vol.~26, no.~4, pp. 984--996, 2014.

\bibitem{biggio2012poisoning}
B.~Biggio, B.~Nelson, and L.~Pavel, ``Poisoning attacks against support vector
  machines,'' in \emph{Proceedings of the 29th International Conference on
  Machine Learning}, 2012.

\bibitem{biggio2011support}
B.~Biggio, B.~Nelson, and P.~Laskov, ``Support vector machines under
  adversarial label noise.'' in \emph{ACML}, 2011, pp. 97--112.

\end{thebibliography}
%

\bibliographystyle{IEEEtran}


\end{document}